\documentclass[camera,letterpaper,nomarginnotes,nonarrowgutter]{jpaper}

\usepackage{cite}
\usepackage{amsmath,amssymb,amsfonts}
\usepackage{algorithmic}
\usepackage{graphicx}
\usepackage{textcomp}
\usepackage{xcolor}
\usepackage{fancyhdr}
\usepackage{hyperref}
\usepackage{titlesec}

\usepackage{glossaries} %
\setkeys{glslink}{hyper=false} %

\usepackage{duckuments} %
\usepackage{balance}
\usepackage{siunitx}
\usepackage{multirow}
\usepackage{tabularx}

\usepackage{booktabs}

\usepackage{setspace}

\newcommand{\X}[0]{{{HARE}}}

\newcommand{\Xone}[0]{{{\X{}-Direct}}}
\newcommand{\Xtwo}[0]{{{\X{}-Filter}}}
\newcommand{\fastHammer}[0]{{\emph{fastHammer}}}
\newcommand{\rh}[0]{RowHammer}

\newcommand{\mitigatingRowHammerAllCitations}[0]{\cite{AppleRefInc, rh-hp,rh-lenovo,greenfield2012throttling, kim2014flipping, kim2014architectural, bains14d, bains14c, bains14a, bains14b, aweke2016anvil, bains2015row, bains2016row, bains2016distributed, son2017making, seyedzadeh2018cbt,irazoqui2016mascat, you2019mrloc, lee2019twice, park2020graphene, yaglikci2021security, yaglikci2021blockhammer, frigo2020trrespass, kang2020cattwo, hassan2021utrr, qureshi2022hydra, saileshwar2022randomized, brasser2017can, konoth2018zebram, van2018guardion, vig2018rapid,  kim2022mithril, lee2021cryoguard, marazzi2022protrr, zhang2022softtrr, joardar2022learning, juffinger2023csi, yaglikci2022hira, saxena2022aqua, enomoto2022efficient, manzhosov2022revisiting, ajorpaz2022evax, naseredini2022alarm, joardar2022machine, hassan2022case, zhang2020leveraging,loughlin2021stop, devaux2021method, fakhrzadehgan2022safeguard, saroiu2022price, loughlin2022moesiprime, han2021surround, mutlu2023fundamentally, woo2023scalable, bock2019riprh, wang2021discreet, bennett2021panopticon, olgun2024abacus, bostanci2024comet, canpolat2025chronus,bostanci2025understanding, yaglikci2024spatial,vittal2025mopac,  qureshi2025autorfm, taneja2026mirza, wu2026apt, lin2025cnc}}

\newcommand{\rowHammerGetsWorseCitations}[0]{\cite{kim2014flipping, kim2020revisiting, frigo2020trrespass, yaglikci2022understanding, orosa2021deeper, mutlu2017rowhammer, mutlu2018rowhammer, mutlu2019rowhammer, mutlu2023fundamentally}}

\newcommand{\nrh}[0]{N_{RH}}
\newcommand{\npr}[0]{N_{PR}}

\newcommand{\pprevref}[0]{p_{REF}}
\newcommand{\pnorefinaburst}[0]{p_{B}}
\newcommand{\cboost}[0]{c_B}
\newcommand{\creduc}[0]{c_R}
\newcommand{\pmin}[0]{p_{min}}
\newcommand{\pmax}[0]{p_{max}}

\newcommand{\pnorefuntilxy}[0]{P_{x,y}}
\newcommand{\pij}[0]{p_{i,j}}
\newcommand{\psa}[0]{p_{sa}}
\newcommand{\pfa}[0]{p_{fa}}
\newcommand{\pattack}[0]{p_{attack}}

\newcommand{\psecure}[0]{p_{security}}
\newcommand{\timeout}[0]{t_{to}}

\newcommand{\trcd}[0]{t_{RCD}}
\newcommand{\tras}[0]{t_{RAS}}
\newcommand{\trp}[0]{t_{RP}}
\newcommand{\trc}[0]{t_{RC}}
\newcommand{\trefi}[0]{t_{REFI}}
\newcommand{\trefw}[0]{t_{REFW}}
\newcommand{\trfc}[0]{t_{RFC}}
\newcommand{\tprofi}[0]{t_{PROFI}}
\newcommand{\tprofw}[0]{t_{PROFW}}
\newcommand{\trrd}[0]{t_{RRD}}

\newcommand{\act}[0]{ACT}
\newcommand{\pre}[0]{PRE}
\newcommand{\refresh}[0]{REF}
\newcommand{\writecmd}[0]{WR}
\newcommand{\rd}[0]{RD}

\newacronym{iqr}{$IQR$}{inter-quartile range}
\newacronym{act}{$\act{}$}{activate}
\newacronym{pre}{$\pre{}$}{precharge}
\newacronym{ref}{$\refresh{}$}{refresh}
\newacronym{wr}{$\writecmd{}$}{write}
\newacronym{rd}{$\rd{}$}{read}
\newacronym{nrh}{$\nrh$}{RowHammer threshold}
\newacronym{npr}{$\npr$}{preventive refresh threshold}
\newacronym{pprevref}{$\pprevref{}$}{preventive refresh probability}
\newacronym{pnorefinaburst}{$\pnorefinaburst$}{the probability of experiencing no preventive refresh during a burst}
\newacronym{pnorefuntilxy}{$\pnorefuntilxy$}{the probability of experiencing \emph{no} preventive refresh until a given $x_{th}$ row activation in $y_{th}$ burst of activations}
\newacronym{cboost}{$\cboost{}$}{the coefficient of boosting preventive refresh probability}
\newacronym{creduc}{$\creduc{}$}{the coefficient of reducing preventive refresh probability}
\newacronym{pij}{$\pij{}$}{the probability of experiencing no preventive refresh after the row activation $i$ at a given batch $j$}
\newacronym{psa}{$\psa{}$}{the probability of a successful attempt}
\newacronym{pfa}{$\pfa{}$}{the probability of a failed attempt}
\newacronym{pattack}{$\pattack{}$}{the probability of a successful attack}
\newacronym{psecure}{$\psecure{}$}{\gls{mech}'s security guarantee over a time period, T,}
\newacronym{trefw}{$\trefw$}{refresh window}
\newacronym{trcd}{$\trcd$}{{trcd definition here}}
\newacronym{tras}{$\tras$}{the latency of fully restoring a DRAM cell's charge}
\newacronym{trp}{$\trp$}{{trp definition here}}
\newacronym{trc}{$\trc$}{the minimum time needed between two consecutive row activations targeting the same bank}
\newacronym{trefi}{$\trefi$}{refresh interval}
\newacronym{trfc}{$\trfc$}{refresh latency}
\newacronym{tprofw}{$\tprofw$}{profiling window}
\newacronym{tprofi}{$\tprofi$}{profiling interval}
\newacronym{trrd}{$\trrd$}{the minimum time needed between two consecutive row activations targeting the same rank}
\newacronym{timeout}{$\timeout$}{timeout period}
\newacronym{pmin}{$\pmin$}{{DEFINE PMIN PLEASE}}
\newacronym{pmax}{$\pmax$}{{DEFINE PMAX PLEASE}}

\newcommand{\rfmth}[0]{T_{RFM}}
\newcommand{\rfmab}[0]{RFM_{ab}}
\newcommand{\rfmsb}[0]{RFM_{sb}}
\newcommand{\aboth}[0]{N_{BO}}
\newcommand{\taboact}[0]{t_{ABO_{ACT}}}
\newcommand{\tbodelay}[0]{t_{BackOffDelay}}

\newcommand{\pmitigth}[0]{T_{FRRFM}}
\newcommand{\pmitig}[0]{FR-RFM}

\newacronym{rfm}{RFM}{refresh management}
\newacronym{prfm}{PRFM}{Periodic RFM}
\newacronym{rfmth}{$\rfmth{}$}{\emph{bank activation threshold}}
\newacronym{rfmab}{$\rfmab{}$}{all bank refresh management}
\newacronym{rfmsb}{$\rfmsb{}$}{same bank refresh management}
\newacronym{prac}{PRAC}{Per Row Activation Counting}
\newacronym{abo}{$ABO$}{Alert Back-Off}
\newacronym{aboth}{$\aboth{}$}{the back-off threshold}
\newacronym{taboact}{$\taboact{}$}{\emph{the window of normal traffic}}
\newacronym{tbodelay}{$\tbodelay{}$}{the delay until a new back-off can be initiated}
\newacronym{jedec}{JEDEC}{Joint Electron Device Engineering Council}

\newacronym{pmitigth}{$\pmitigth{}$}{the RFM period of \pmitig{}}
\newacronym{pmitig}{\pmitig}{Fixed-Rate RFM}

\newcommand{\cots}[0]{COTS}
\newcommand{\comra}[0]{CoMRA}
\newcommand{\simra}[0]{SiMRA}

\newacronym{cots}{\cots{}}{commercial off-the-shelf}
\newacronym{comra}{\comra{}}{}
\newacronym{comraL}{which we call \comra{}}{\underline{co}nsecutive \underline{m}ultiple-\underline{r}ow \underline{a}ctivation}
\newacronym{simraL}{which we call \simra{}}{\underline{si}multaneous \underline{m}ultiple-\underline{r}ow \underline{a}ctivation}
\newacronym{simra}{\simra{}}{}

\newcommand*\nCHIPS{180}
\newcommand*\nMODULES{25}

\newcommand{\secref}[1]{§\ref{#1}}
\newcommand{\figref}[1]{Fig.~\ref{#1}}

\usepackage{tikz}

\newcommand*\circled[1]{\tikz[baseline=(char.base)]{\node[shape=circle,fill,inner sep=0.5pt] (char) {\textcolor{white}{#1}};}}

\newcommand{\head}[1]{{\noindent\textbf{#1.} }}

\definecolor{nbs}{rgb}{0.88, 0.07, 0.37}
\definecolor{iy}{rgb}{0.0, 0.36, 0.05}
\definecolor{gfored}{rgb}{0.580, 0.050, 0.211}
\definecolor{bluepigment}{rgb}{0.2, 0.2, 0.6}
\definecolor{brandeisblue}{rgb}{0.0, 0.44, 1.0}
\definecolor{ouscolor}{rgb}{0.0, 0.2, 0.4}
\definecolor{omc}{rgb}{1, 0.01, 0.01}

\definecolor{ra}{rgb}{0.24, 0.44, 0.88} 
\definecolor{rd}{rgb}{0.86, 0.15, 0.50} 
\definecolor{rc}{rgb}{0.83, 0.31, 0.00} 
\definecolor{rb}{rgb}{0.47, 0.37, 0.94} 
\definecolor{re}{rgb}{0.58, 0.40, 0.00} 
\definecolor{cq}{rgb}{0.10, 0.55, 0.30} 

\definecolor{burgundy}{rgb}{0.5, 0.0, 0.13}

\definecolor{hs}{rgb}{0.059, 0.322, 0.729}

\definecolor{vetc}{rgb}{0.0, 0.5, 0.5}

\newif\ifdraft
\draftfalse

\newif\ifrevision
\revisionfalse

\ifrevision
    \newcommand{\revcommon}[1]{\textcolor{cq}{#1}}
    \newcommand{\reva}[1]{\textcolor{ra}{#1}}
    
    \newcommand{\revc}[1]{\textcolor{rc}{#1}}
    \newcommand{\revd}[1]{\textcolor{rd}{#1}}
    \newcommand{\reve}[1]{\textcolor{re}{#1}}

    \newcommand\revmid[2]{\todo[linecolor=#1,backgroundcolor=#1!15,bordercolor=#1]{\textcolor{black}{\textbf{#2}}}}

    \newcommand{\labela}[1]{\revmid{ra}{A#1}}
    
    \newcommand{\labelc}[1]{\revmid{rc}{C#1}}
    \newcommand{\labeld}[1]{\revmid{rd}{D#1}}
    \newcommand{\labele}[1]{\revmid{re}{E#1}}
    \newcommand{\labelcq}[1]{\revmid{cq}{CQ#1}}

    \definecolor{babyblueeyes}{rgb}{0.63, 0.79, 0.95} 
    \usepackage[colorinlistoftodos,prependcaption,textsize=small]{todonotes}
    
\else
    \newcommand{\revcommon}[1]{{#1}}
    \newcommand{\reva}[1]{{#1}}
    
    \newcommand{\revc}[1]{{#1}}
    \newcommand{\revd}[1]{{#1}}
    \newcommand{\reve}[1]{{#1}}

    \newcommand\revmid[2]{}

    \newcommand{\labela}[1]{\revmid{ra}{A#1}}
    
    \newcommand{\labelc}[1]{\revmid{rc}{C#1}}
    \newcommand{\labeld}[1]{\revmid{rd}{D#1}}
    \newcommand{\labele}[1]{\revmid{re}{E#1}}
    \newcommand{\labelcq}[1]{\revmid{cq}{CQ#1}}

    \definecolor{babyblueeyes}{rgb}{0.63, 0.79, 0.95} 
\fi

\ifdraft
    \usepackage[colorinlistoftodos,prependcaption,textsize=scriptsize]{todonotes} 
    
    \paperwidth=\dimexpr \paperwidth + 4cm\relax
    \oddsidemargin=\dimexpr\oddsidemargin + 2cm\relax
    \evensidemargin=\dimexpr\evensidemargin + 2cm\relax
    \marginparwidth=\dimexpr \marginparwidth + 2cm\relax
    
    \newcommand{\nbcomment}[1]{\todo[size=\scriptsize, linecolor=nbs, bordercolor=nbs, backgroundcolor=nbs!5]{\textcolor{nbs}{\textbf{@nb:} #1}}}
    \newcommand\param[1]{{\color{blue}{#1}}}
    \newcommand\nb[1]{{\color{nbs}{#1}}}

    \newcommand{\atbcomment}[1]{\todo[size=\scriptsize, linecolor=orange, bordercolor=orange, backgroundcolor=white]{\textcolor{orange}{\textbf{@atb:} #1}}}
    \newcommand\atb[1]{{\color{orange}{#1}}}
    
    \newcommand{\ouscomment}[1]{\todo[size=\scriptsize, linecolor=orange, bordercolor=orange, backgroundcolor=white]{\textcolor{ouscolor}{\textbf{@ous:} #1}}}

    \newcommand{\iey}[1]{\textcolor{iy}{#1}}
    \newcommand{\ieycomment}[1]{\todo[size=\scriptsize, linecolor=orange, bordercolor=orange, backgroundcolor=white]{\textcolor{iy}{\textbf{@iey:} #1}}}

    \newcommand{\agycomment}[1]{\todo[size=\scriptsize, linecolor=purple, bordercolor=purple, backgroundcolor=white]{\textcolor{purple}{\textbf{@agy:} #1}}}

    \newcommand{\hscomment}[1]{\todo[size=\scriptsize, linecolor=orange, bordercolor=orange, backgroundcolor=white]{\textcolor{hs}{\textbf{@hs:} #1}}}
    
    \newcommand{\outline}[1]{\textcolor{red}{\texttt{\textbf{Outline:}~#1}}}
    \newcommand{\question}[1]{\textcolor{blue}{\texttt{\textbf{Question:}~#1}}}
    \newcommand{\inlinetodo}[1]{\textcolor{red}{\textbf{TODO:} #1}}

    \newcommand{\vet}[1]{\todo[size=\scriptsize, linecolor=vetc, bordercolor=vetc, backgroundcolor=vetc!10]{\textcolor{vetc}{\textbf{VET FROM HERE.} #1}}}
\else
    
    \definecolor{shadecolor}{RGB}{180,180,180}
    \definecolor{darkgray}{rgb}{0.86, 0.86, 0.86}

    \newcommand\param[1]{{\color{black}{#1}}}

    \newcommand{\nbcomment}[1]{}
    \newcommand\nb[1]{#1}

    \newcommand\atb[1]{#1}
    \newcommand{\atbcomment}[1]{}

    \newcommand{\ouscomment}[1]{}
    
    \newcommand{\iey}[1]{#1}
    \newcommand{\ieycomment}[1]{}

    \newcommand{\agycomment}[1]{}

    \newcommand{\hscomment}[1]{}

    \newcommand{\outline}[1]{}
    \newcommand{\question}[1]{}
    \newcommand{\inlinetodo}[1]{}
    \newcommand{\vet}[1]{}
\fi

\definecolor{frenchblue}{rgb}{0.19, 0.55, 0.91}

\definecolor{ibmblue}{RGB}{100,143,255}
\definecolor{ibmyellow}{RGB}{255,176,0}

\usepackage[most]{tcolorbox} 
\tcbset{before skip=4.1pt, after skip=4.1pt}

\newtcolorbox[auto counter]{obsx}[3][]{%
    colframe = #2!45,
    colback  = #2!10,
    coltitle = #2!20!black, 
    colbacktitle=#2!20,
    coltitle=black,
    fonttitle=\bfseries, 
    title=#3~\thetcbcounter.\ ,
    enhanced,
    attach boxed title to top left={yshift=-2.8mm, xshift=0.15cm},
    bottom=-2.2pt,
    #1%
}

\usepackage[most]{tcolorbox} 
\newtcolorbox[auto counter]{tkx}[2][]{%
    enhanced, breakable, center title,
    colframe = #2!45,
    colback  = #2!10,
    colbacktitle=#2!20,
    left=-0.5pt,
    right=-0.5pt,
    bottom=-2pt,
    top=-0.25pt,
    #1%
}
\newcounter{obs}
\newcommand\observation[1]{
\refstepcounter{obs}
\begin{tkx}{ibmyellow}
\noindent\textbf{Observation~\theobs.} #1
\end{tkx}
}

\newcounter{tkw}
\newcounter{hypo}
\newcounter{opprt}
\newcommand\opportunity[1]{
\refstepcounter{opprt}
\begin{tkx}{ibmblue}
\noindent\textbf{Opportunity~\theopprt.} #1
\end{tkx}
}

\begin{document}

\title{\X{}: Fast Read Disturbance Threshold Profiling\\in Modern DRAM Chips via Reduced Timing Constraints}

\author{F. Nisa Bostanc{\i}
\quad \.{I}smail E. Y\"{u}ksel 
\quad Ataberk Olgun
\quad Harsh Songara \\
Andreas Kosmas Kakolyris 
\quad Haocong Luo
\quad Konstantinos Marios Sgouras
\quad Onur Mutlu
\vspace{-4mm} \\\\
{SAFARI Research Group}
\\
ETH Zurich}

\fancyhead{}
\renewcommand{\headrulewidth}{0pt}

\maketitle


\setstretch{0.97}
\begin{abstract}
{Modern DRAM chips are subject to read disturbance errors.
These errors manifest as security-critical bitflips in a victim DRAM row physically nearby {an aggressor row that is} repeatedly activated (opened) (RowHammer) or kept open for a long time (RowPress).
{To prevent read disturbance bitflips securely and with low overhead, s}tate-of-the-art read disturbance mitigations rely on accurate and exhaustive profiling of the read disturbance threshold (RDT) (i.e., the \nb{minimum} 
number of aggressor row activations needed to induce the first bitflip)}.
\nb{However}, \nb{RDT is \emph{not} static:} it changes significantly and unpredictably over time, motivating \emph{online} profiling mechanisms that dynamically adjust a mitigation mechanism's RDT configuration.
Unfortunately, exhaustively profiling a DRAM module's RDT is costly
for such online profilers
\nb{(e.g., \param{hundreds of} seconds for \param{a 16-bank DDR4 module}) for two main reasons.}
First, measuring the RDT of a single DRAM row requires activating its aggressor rows many times (e.g., \param{4K} times for DDR4 modules) before a bitflip is observed.
Second, \nb{due to manufacturing process variation, each DRAM row can have a different RDT, requiring many rows to be profiled (e.g.,} \param{65'536} rows \param{per bank}).

Our goal is to design a fast \nb{online} RDT
profiling technique that identifies the 
minimum RDT across a DRAM module at low cost.
To this end, we introduce \X{}, 
an \nb{online} RDT profiling technique that intentionally exacerbates read disturbance to 
quickly profile a DRAM module. \X{}'s key idea is to reduce the timing 
constraints between consecutive DRAM commands during profiling.
Doing so significantly reduces profiling time by 1)~directly reducing the 
execution time of the profiling command sequence and 2)~exacerbating read 
disturbance such that the first bitflip occurs at a 
smaller activation 
count than with the conventional RowHammer sequence.

From our detailed characterization of \param{\nCHIPS{}} real DDR4 DRAM chips from 
\param{four} major DRAM manufacturers, we make three key observations.
First, \X{} measures the RDT of a DRAM row \param{31.2\%} faster 
than the RowHammer profiler and the minimum RDT of a DRAM module, with an average reduction of \param{21.7\%}. Second, \X{} flips more than 95\% of the cells that the RowHammer access pattern flips.
Third, \X{} 
{preserves the RDT ranking of rows well enough to}
exhaustively reprofile a small set of rows to measure the minimum \rh{} RDT.
Our end-to-end evaluations show that \X{} enables significant system performance improvements and outperforms the RowHammer profiler.

\end{abstract}

\setstretch{1}
\section{Introduction}
\label{sec:introduction}

Memory isolation is a fundamental property for system robustness (i.e.,
reliability, security, and safety). Accesses to one memory location should
\emph{not} induce unintended side-effects on data stored in other
(\emph{unaccessed}) memory locations. Unfortunately, modern DRAM chips are
subject to read disturbance, where accessing a DRAM row induces bitflips in
physically nearby DRAM rows that are \emph{not} accessed, breaking memory
isolation. RowHammer~\cite{kim2014flipping} and RowPress~\cite{luo2023rowpress}
are two prominent read disturbance phenomena, in which a victim DRAM row
experiences bitflips when a nearby aggressor row is repeatedly activated
(i.e., hammered) or kept open for a long period of time (i.e., pressed), respectively.

Read disturbance worsens as DRAM manufacturing technology scales down to
smaller node sizes~\cite{kim2020revisiting,orosa2021deeper,luo2023rowpress}.
For example, the number of aggressor row activations needed to induce the first
bitflip has reduced by more than an order of magnitude in less than a
decade~\cite{kim2020revisiting}. Therefore, a large body of work proposes 
read disturbance mitigation mechanisms. To prevent read disturbance bitflips
securely and with low overhead, many of these read disturbance mitigation
mechanisms
\mitigatingRowHammerAllCitations{} 
rely on accurate knowledge of the read disturbance threshold (RDT), i.e., the
number of aggressor row activations needed to induce the first bitflip in a
victim row. A row's RDT is typically profiled by activating its 
aggressor rows repeatedly until a bitflip is observed and determining the actual hammer count with a binary search approach. 
The mitigation mechanisms are configured for an RDT value (e.g., minimum profiled 
RDT value, or the worst-case RDT value) and 
perform preventive actions (e.g., refreshing victim rows) before a row's activation
count reaches the RDT value.

Unfortunately, the RDT of a DRAM row is \emph{not} static; it changes
significantly and unpredictably over time~\cite{olgun2025variable}. Thus, a row can
experience bitflips at lower activation counts compared to the previously 
profiled RDT value, resulting in any mitigation configured for that RDT to fail 
preventing attacks. To mitigate bitflips in the presence of variable read disturbance effect, 
mitigation mechanisms require dynamic adjustments to the RDT through online RDT profiling.

RDT profiling is prohibitively slow for two main
reasons. First, measuring the RDT of a single DRAM row requires activating its
aggressor rows many times (e.g., at least \param{4}K times for DDR4
chips (\secref{sec:characterization})) before a bitflip is observed. Second, because
of manufacturing process variation, each DRAM row can have a different 
RDT (i.e., can fail at different aggressor row activation counts), requiring 
many rows to be profiled.
Thus, profiling the RDT of many rows (e.g., 65'536 rows per DRAM bank in 
a 16-bank DDR4 module~\cite{jedec2017ddr4}) takes significant time (e.g., hundreds of seconds~\cite{olgun2025variable}).

Our goal is to design a fast RDT profiling technique that identifies the
minimum RDT of a DRAM module at low cost. To this end, we
introduce \X{}, a fast RDT profiling technique that intentionally exacerbates 
read disturbance to quickly profile a DRAM module. \X{}'s key idea is to 
reduce the timing constraints between consecutive DRAM commands in the 
profiling sequence. Doing so reduces the profiling time significantly by 
1)~reducing the time it takes to \nb{perform one hammer}, and 2)~exacerbating the read disturbance to 
induce bitflips with far fewer hammers than conventional RowHammer patterns.

In this paper, we present the first analysis of read disturbance under 
reduced timing parameters as a fast RDT profiling technique, on \param{\nCHIPS{}} 
commercial off-the-shelf (COTS) DDR4 DRAM chips from \param{four} major DRAM
manufacturers (in \param{\nMODULES{}} DRAM modules). 
\iey{Based on our characterization, we make \param{three} key observations that form the basis of \X{}'s design.}

First, hammering a row with reduced timing constraints, which we call \fastHammer{}, reduces the row's RDT in all tested real DRAM chips from all four major manufacturers \nb{(by 27.0\%-33.5\% depending on the manufacturer)}. \iey{Thus, \fastHammer{} induces the first bitflip with fewer aggressor row activations than the conventional RowHammer pattern. Second, \fastHammer{} induces bitflips in more than \param{95\%} of the cells that the conventional RowHammer pattern flips, at every tested configuration. Therefore, \fastHammer{} covers \emph{nearly all} the cells that RowHammer disturbs. Third, \fastHammer{} preserves the ordering of rows by RDT: in \emph{all} \param{\nMODULES{}} tested modules, the RowHammer-weakest row falls within the top \param{5.6\%} of rows ranked by their \fastHammer{} RDT. Therefore, a module's minimum RowHammer RDT can be identified by exhaustively profiling \emph{only} a small candidate set of rows.}

\iey{\X{} builds on these observations in two variants. \Xone{} profiles every row in the module with \fastHammer{} and reports the lowest measured RDT. Because \fastHammer{} lowers a row's RDT, \Xone{}'s estimate is conservative, and thus, inherently applies a guardband to the mitigation's configured RDT. \Xtwo{} improves the accuracy of this estimate at the cost of additional profiling time: it ranks a batch of rows by their \fastHammer{} RDT, reprofiles \emph{only} the top-ranked rows with the RowHammer pattern, and reports the minimum RowHammer RDT among them. We implement both variants as periodic online profilers that profile all rows of a module within a profiling window and continuously update the RDT configuration of the underlying RowHammer mitigation.}

\iey{We evaluate \X{} against two baselines. To quantify the benefit of online RDT profiling, we compare \X{} to a system with \emph{no} online profiler using a probabilistic error model of read disturbance bitflips under variable read disturbance, built from our characterization results. Across \param{1M} Monte Carlo trials, a system with no online profiler experiences an uncorrectable error (i.e., a read disturbance bitflip pattern that the DRAM chip's on-die ECC cannot correct) in \param{99.9\%} of the trials, whereas \Xone{} reduces this probability to \param{6.2\%} and \Xtwo{} reduces it by up to \param{12.1}$\times$. To quantify the benefit of \fastHammer{}, we compare \X{} to a system with a conventional RowHammer profiler using Ramulator 2.0, a cycle-level DRAM simulator, across
quad-core 
memory-intensive workload mixes drawn from the SPEC CPU2006, SPEC CPU2017, TPC, MediaBench, and YCSB suites. Over \param{36} configurations spanning three state-of-the-art RowHammer mitigations (i.e., PRAC~\cite{jedec2024jesd795c}, Graphene~\cite{park2020graphene}, and PARA~\cite{kim2014flipping}), three RDT values, and four profiling intervals, \Xone{} and \Xtwo{} improve geomean weighted speedup over the conventional RowHammer profiler in \emph{every} configuration, by \param{0.9\%} to \param{54.4\%} and \param{1.5\%} to \param{47.5\%}, respectively.}

\iey{This paper makes the following key contributions:}
\begin{itemize}
\item We present the first characterization of read disturbance under reduced DRAM timing constraints as a method for fast RDT profiling, using \param{\nCHIPS{}} real DDR4 DRAM chips (\param{\nMODULES{}} modules) from \param{four} major DRAM manufacturers.

\item We show that \fastHammer{} induces bitflips in more than \param{95\%} of the cells that the conventional \rh{} access pattern flips, and that it preserves the ordering of rows by RDT \reva{well enough}\labela{3}\label{resp:a3_3}: the \rh{}-weakest row of \emph{every} tested module falls within the top \param{5.6\%} of \fastHammer{}-ranked rows. A module's minimum \rh{} RDT can therefore be identified by exhaustively profiling only a small candidate set of rows.

\item We propose \X{}, the first online RDT profiling technique that intentionally exacerbates read disturbance to profile a DRAM module quickly, in two variants: \Xone{}, which reports a conservative minimum RDT, and \Xtwo{}, which recovers the conventional \rh{} RDT by reprofiling only the top-ranked rows.

\item We develop a probabilistic error model for read disturbance bitflips under variable read disturbance, parameterized by our characterization data, and use it to show that a mitigation configured by one-time profiling \emph{cannot} operate reliably: it exhibits an uncorrectable error in \param{99.9\%} of simulated systems, with a mean time to failure of only \param{8.5} minutes, and reaching a negligible error probability requires halving the configured RDT.

\item We show that \X{} substantially improves both system reliability and performance: it reduces the probability of an uncorrectable error from \param{66.8\%} to \param{0.67\%} compared to a system with \emph{no} online profiler at the same \param{20\%} guardband, and outperforms a conventional \rh{} profiler in \emph{every} evaluated configuration, by up to \param{54.4\%}.
\end{itemize}

\section{Background}
\begin{figure}[ht]
\centering
\includegraphics[width=0.87\linewidth]{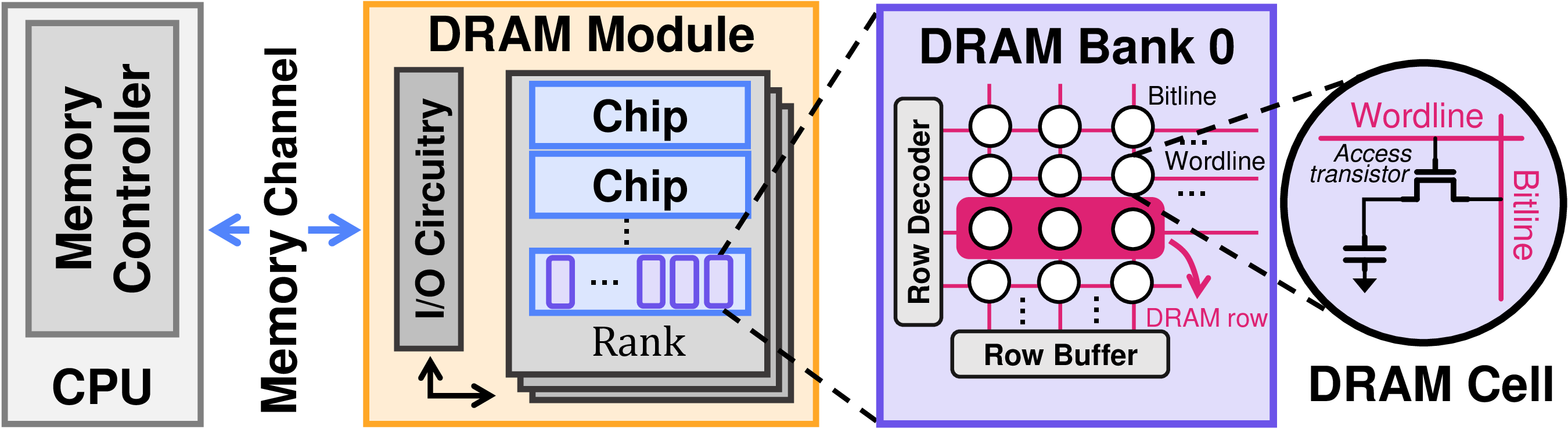}
\caption{DRAM organization}
\label{fig:dram_organization}
\end{figure}

\head{DRAM Organization}
\label{sec:background}
Figure~\ref{fig:dram_organization} shows DRAM-based main memory organization.
In a CPU, a memory controller interfaces with a DRAM channel (containing DRAM module(s)) to serve memory requests.
Each module contains multiple DRAM ranks (i.e., a number of DRAM chips operating in lockstep), with each DRAM chip comprising multiple DRAM banks.
A DRAM bank comprises a two-dimensional array of DRAM cells organized into columns and rows. 
Each DRAM cell stores a single bit of information as charge stored in a capacitor. 
To access the cell, an access transistor controlled by a wordline 
connects the cell to a bitline. 
When the wordline is asserted, the access transistors connect all cells of a row to their corresponding bitline, perturbing their voltage.
A structure called the \textit{row buffer}, consisting of an array of sense amplifiers, senses and amplifies this voltage perturbation.

\head{DRAM Operation}
To access data, the memory controller first issues an activate (\texttt{ACT}) command, which ``\textit{opens}'' the corresponding DRAM row and then can read or write data. 
Before accessing another row in the same bank, the memory controller must first ``\textit{close}'' the currently open row by issuing a precharge (\texttt{PRE}) command.
The memory controller schedules DRAM commands according to timing constraints~\cite{kim2012case,lee2013tiered,lee2015adaptive}, such as $t_{RAS}$ (\iey{i.e., the minimum time between opening a row with an \act{} command and closing the row with a \pre{} command)}, $t_{RP}$~\iey{(i.e., the minimum time between sending a \pre{} command and opening a row with an \act{} command)} and $t_{RC}$ (i.e., the minimum delay between two consecutive \texttt{ACT} commands to the same bank).

\head{DRAM Read Disturbance}
\label{sub:dram_read_disturbance}
\rh{} and RowPress occur when a row is (i) repeatedly opened and closed, i.e., hammered, or (ii) kept open for a prolonged period, i.e., pressed, causing bitflips in adjacent rows.
The number of times the ``\textit{aggressor}'' row must be hammered or pressed to induce bitflips in adjacent ``\textit{victim}'' rows is called the \textit{read disturbance threshold} (RDT).
DRAM technology node scaling has increased the severity of read disturbance phenomena~\rowHammerGetsWorseCitations{}.
Many prior works~\mitigatingRowHammerAllCitations{} have proposed various mitigation techniques that prevent read disturbance bitflips, particularly those caused by \rh{}.
To ensure robust DRAM operation, all of these techniques require accurate knowledge of the RDT.

\section{Motivation and Goal}
\begin{figure}[ht]
\centering
\includegraphics[width=\linewidth]{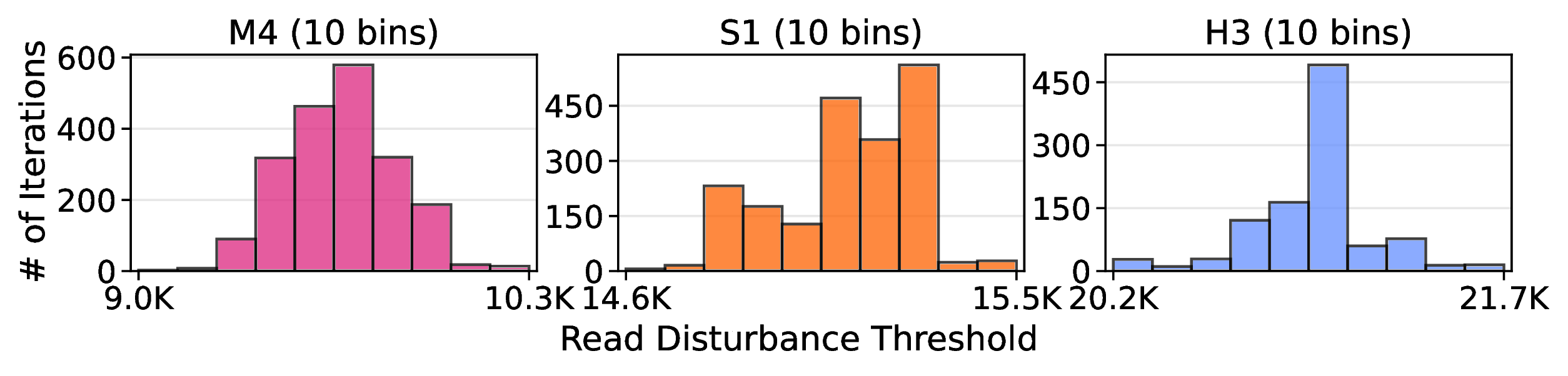}
\caption{\revcommon{RDT value histogram for a single victim row in each tested module.}}
\label{fig:row-vrd}
\end{figure}

\head{\revcommon{Variable Read Disturbance}}
\label{resp:cq1}
RDT of a DRAM row is \emph{not} static: it changes
significantly and unpredictably over time~\cite{olgun2025variable}. Thus, a row can
experience bitflips at lower activation counts compared to the previously 
profiled RDT value, 
\iey{causing any mitigation configured for that RDT to fail to prevent bitflips.} \labelcq{1} \revcommon{\figref{fig:row-vrd} shows the change in a single victim row's RDT in three representative modules
across profiling iterations.\footnote{\secref{sec:methodology} describes our real DRAM testing infrastructure and methodology.} Each iteration measures the row's RDT once by executing the profiling routine. Each subplot depicts the histogram of a victim row's RDT in one module, separated into 10 bins. We observe that each row's RDT significantly changes across repeated measurements, similar to~\cite{olgun2025variable}.} 

\begin{figure}[ht]
\centering
\includegraphics[width=\linewidth]{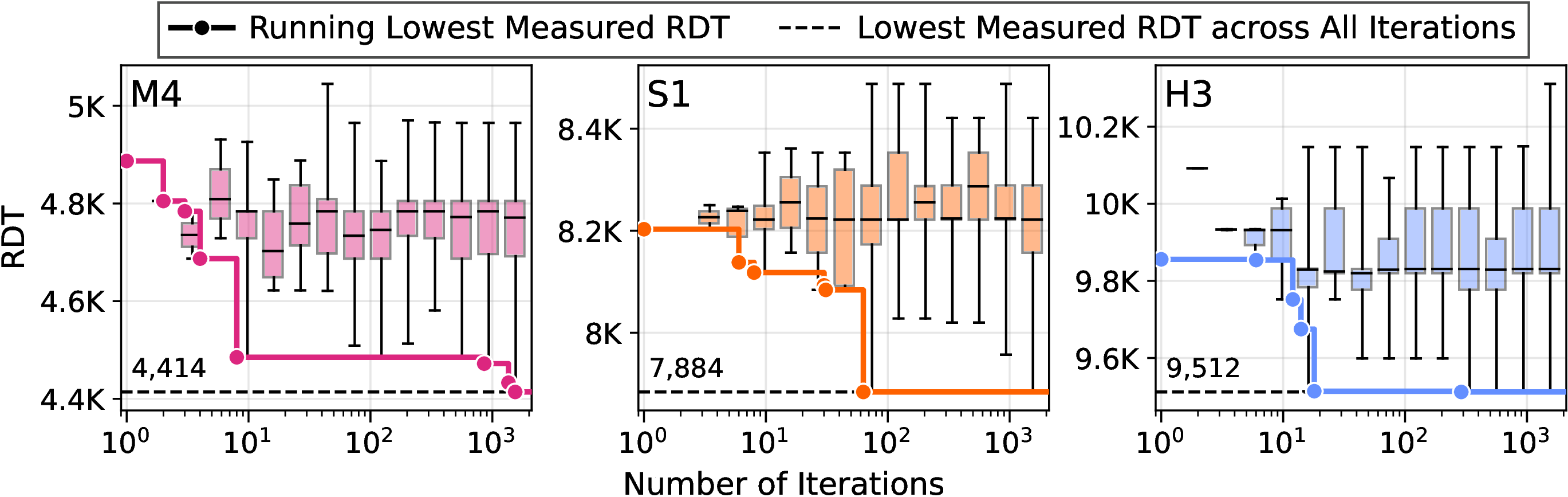}
\caption{\revcommon{Minimum RDT of a module across three representative modules tested for 2K profiling iterations.}}
\label{fig:module-vrd}
\end{figure}

A RowHammer mitigation is typically configured for a minimum RDT across all rows in a module. \revcommon{We analyze how the minimum RDT changes across profiling iterations in three representative modules from three manufacturers. In each profiling iteration, we measure the RDT of 3K rows per module.~\figref{fig:module-vrd} plots our measurements across 2K iterations. Each box represents a bin of 200 iterations (whiskers: the highest and the lowest minimum RDT in the bin). The solid line depicts the running minimum RDT of the module. The dashed line depicts the minimum RDT across 2K iterations.}

\revcommon{We observe that the minimum RDT changes across iterations and can be measured as late as the 1'539th iteration out of 2K iterations (in M4). We conclude that the minimum RDT across rows varies significantly across iterations.}

\begin{figure}[ht]
\centering
\includegraphics[width=\linewidth]{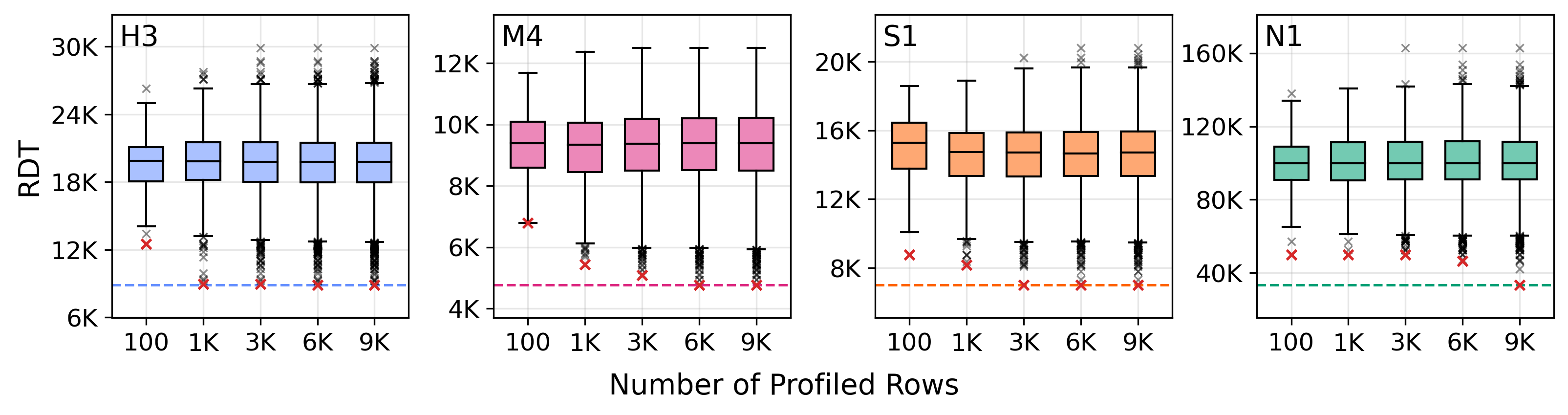}
\caption{\reva{Minimum RDT of a DRAM row group with increasing size}}
\label{fig:row-sweep}
\end{figure}

\head{\reva{Spatial Variation in RDT across DRAM Rows}}
\label{sec:motive-spatial}
\reva{Prior works~\cite{yaglikci2024spatial,orosa2021deeper} show that RDT varies significantly and irregularly across DRAM rows.~\figref{fig:row-sweep} shows the RDT distributions of row groups of increasing size, from 100 to 9K rows, where each group includes the rows of smaller groups. The red mark shows the minimum RDT measured in the selected row group and the dashed line shows the minimum RDT measured across all groups. We observe that the minimum RDT does \emph{not} stabilize at a size under 9K across all tested modules.}

\head{RDT Profiling is Time Consuming}
\label{subsec:profiling_cost}
    To mitigate bitflips in the presence of VRD, 
RDT profiling is prohibitively slow for two main
reasons. First, measuring the RDT of a single row requires activating its
aggressor rows many times (e.g., at least \param{4}K times for DDR4
chips~\cite{olgun2025variable}) before a bitflip is observed. 
\iey{Second, profiling one row is not enough. \nb{Due to m}anufacturing process variation each row \nb{can} have a different RDT~\cite{yaglikci2024spatial}. A \rh{} mitigation is typically configured for the \emph{minimum} RDT across the module (including those
that are used and standardized by industry, e.g., PRAC~\cite{jedec2024jesd795c,canpolat2024understanding}), so a profiler must visit \emph{every} row (e.g., \param{65'536} rows per bank in a 16-bank DDR4 module). Thus, profiling all rows of a DRAM module takes \param{hundreds} of~seconds~\cite{olgun2025variable}.}
\nb{This time is prohibitively long for an online profiler because (i)~at the time of RDT profiling, DRAM bandwidth cannot be utilized for demand memory requests, and (ii)~enabling reliable operation with low performance overhead requires frequent profiling (as we show in~\secref{sec:security_analysis}).}
\iey{\noindent{Our goal} is to design a fast online RDT profiling technique that identifies the minimum RDT of a DRAM module at low cost.}

\section{COTS DRAM Testing Methodology}
\label{sec:methodology}

\subsection{COTS DRAM Testing Infrastructure}
\label{subsec:infra}
\begin{figure}[ht]
\centering
\includegraphics[width=0.87\linewidth]{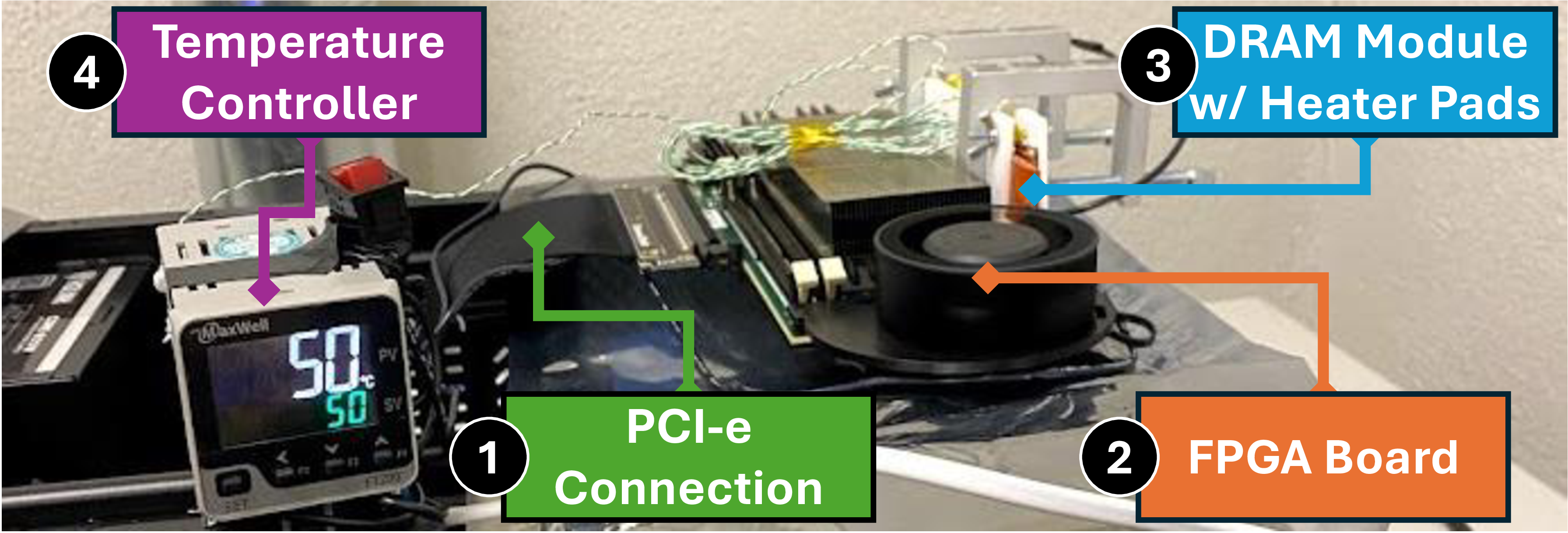}
\caption{Our DRAM Bender~\cite{olgun2023dram} based experimental setup.}
\label{fig:infra}
\end{figure}

We conduct COTS DRAM chip experiments using DRAM Bender~\cite{safari-drambender,olgun2023dram}, an FPGA-based DDR4 testing infrastructure that provides precise control of DDR4 commands. \figref{fig:infra} shows our experimental setup that consists of four main components: 1) a host machine that generates the test program, 2) an FPGA board~\cite{alveo_u200}, programmed with DRAM Bender, 3) a temperature sensor and heater pads to maintain a target temperature, and 4) a temperature controller~\cite{maxwellFT200} that keeps the temperature at the desired level. 

\noindent\textbf{Eliminating Interference Sources.}
To observe read disturbance bitflips {at} the circuit level,
we eliminate potential sources of interference, by taking \param{three} measures, similar to the methodology used by prior works~\cite{kim2020revisiting, orosa2021deeper, yaglikci2022understanding, hassan2021utrr, luo2023rowpress}.
First, we disable periodic refresh during the execution of test programs to prevent potential on-DRAM-die TRR mechanisms~\cite{frigo2020trrespass, hassan2021utrr} from refreshing victim rows so that we can observe the DRAM chip's behavior at the circuit-level.
Second, we strictly bound the execution time of test programs within the refresh window of the DRAM chips to avoid data retention failures interfering with read disturbance failures.
Third, we {verify} that the modules and chips have neither rank-level nor on-die ECC~\cite{patel2020beer, patel2021harp}.
{With these measures,} we directly observe and analyze all bitflips without interference.

\subsection{COTS DDR4 DRAM Chips Tested}
\label{subsec:tested_chips}

Table~\ref{tab:dram_chips} provides the \nCHIPS{} (\nMODULES{}) COTS DDR4 DRAM chips (modules).
We label each module by its manufacturer and an index, i.e., H1--H6 (SK Hynix), M1--M10 (Micron), S1--S7 (Samsung), and N1--N2 (Nanya).







\begin{scriptsize}
\begin{table}[ht]
\centering
\caption{Summary of DDR4 DRAM chips tested.}
\label{tab:dram_chips}
\resizebox{\columnwidth}{!}{%
\renewcommand{\arraystretch}{0.95}
\begin{tabular}{lccccc}
\textbf{Chip Mfr.} &  \textbf{\#Modules} &  \textbf{\#Chips} & \textbf{Die Rev.} & \textbf{Density} & \textbf{Org.} \\
\hline\hline
            &        3 &      24 &        A &     8Gb &   x8 \\
  SK Hynix         &         1 &       8 &        D &     8Gb &   x8 \\
          &         2 &      16 &        C &    16Gb &   x8 \\
\midrule
   &         2 &      16 &        R &     8Gb &   x8 \\
     Micron      &         4 &      16 &        E &    16Gb &  x16 \\
          &         4 &      32 &        F &    16Gb &   x8 \\
\midrule
   &         1 &       4 &        C &     4Gb &  x16 \\
  Samsung        &         1 &       8 &        E &     4Gb &   x8 \\
          &         5 &      40 &        B &    16Gb &   x8 \\
\midrule
    Nanya &         2 &      16 &        C &     8Gb &   x8 \\
\hline
\end{tabular}
}
\end{table}
\end{scriptsize}

\noindent\textbf{Logical-to-Physical Row Mapping.}
DRAM manufacturers use mapping schemes to translate logical (memory-controller-visible) addresses to physical row addresses~\cite{kim2014flipping, smith1981laser, horiguchi1997redundancy, keeth2001dram, itoh2013vlsi, liu2013experimental,seshadri2015gather, khan2016parbor, khan2017detecting, lee2017design, tatar2018defeating, barenghi2018software, cojocar2020rowhammer,  patel2020beer,kim2012case,frigo2020trrespass}. To account for in-DRAM row address mapping, we reverse engineer the physical row address layout {in all chips}, following the prior works' methodology~\cite{kim2020revisiting, orosa2021deeper, yaglikci2022understanding, luo2023rowpress}.

\noindent\textbf{Read Disturbance Vulnerability Metric.}
To characterize a DRAM module's read disturbance vulnerability, we use read disturbance threshold (RDT) via a double-sided RowHammer access pattern, where we count the hammer cycles (i.e., each pair of activations to the aggressor rows as one hammer). A {lower} RDT indicates a {higher} read disturbance vulnerability.

\noindent\textbf{RDT Algorithm.}
For every tested parameter we evaluate (e.g., data pattern and temperature), we find the RDT for each tested victim row using the bisection-method algorithm used by prior works~\cite{orosa2021deeper,yaglikci2022understanding,luo2023rowpress}. We terminate the RDT search when the difference between the current and previous RDT measurements is no larger than 1\% of the previous measurements. For every tested row, we repeat the RDT search five times and report the minimum RDT value we observe.

\noindent\textbf{Data Pattern.}
 We use the four data patterns (\texttt{0x00},~\texttt{0xFF},~\texttt{0xAA}, and~\texttt{0x55}) that are widely used in memory reliability testing~\cite{vandegoor2002address,khan2014efficacy} and by prior work on DRAM characterization (e.g.,~\cite{kim2014flipping,kim2020revisiting,orosa2021deeper,luo2023rowpress,yaglikci2024spatial,olgun2025variable,luo2024experimental,tugrul2025understanding,olgun2024read}). We fill aggressor rows with these data patterns while initializing victim rows with the negated data pattern (e.g., if aggressor rows are~\texttt{0x00}, victim rows are~\texttt{0xFF}). For each DRAM row, we define the worst-case data pattern (\emph{WCDP}) as the data pattern that causes the lowest RDT. All experiments are conducted using WCDP.

\noindent\textbf{Temperature.} 
All experiments are conducted at 80$^{\circ}$C unless stated otherwise.

\noindent\textbf{Number of Instances Tested.} To maintain a reasonable testing time, we test \param{3000} rows in one bank per DRAM module: \param{1000} rows each from the beginning, middle, and end of the bank.

\section{Fast RDT Profiling\\via Reduced Timing Constraints}
\label{sec:characterization}

\subsection{Hammering 
with Reduced Timing Constraints}

\begin{figure}[ht]
\centering
\includegraphics[width=0.83\linewidth]{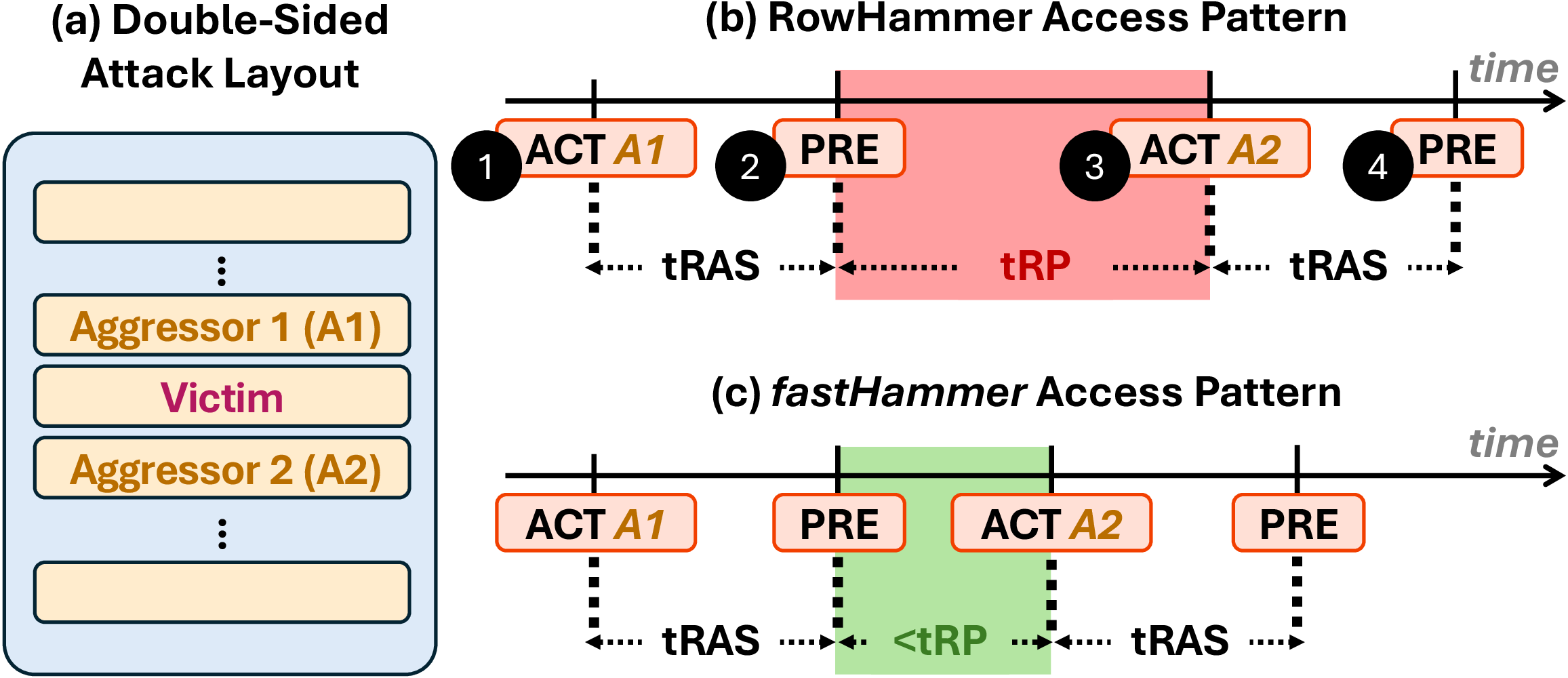}
\caption{Double-sided RowHammer and \fastHammer{} access pattern}
\label{fig:fast_hammer}
\end{figure}

In this work, we construct a fast RowHammer access pattern, \fastHammer{}, that violates the timing constraints between DRAM commands in the hammer sequence. \figref{fig:fast_hammer} shows (a)~the double-sided attack layout in DRAM, (b)~the conventional RowHammer double-sided access pattern and (c)~the \fastHammer{} double-sided access pattern with reduced \gls{pre}-to-\gls{act} latency.
\footnote{We observe that reducing the \gls{act} to \gls{pre} latency does \textit{not} yield a lower average RDT, hence we do \textit{not} consider it as the main method to exacerbate read disturbance for fast profiling.}

\figref{fig:fast_hammer}(a) shows a double-sided attack layout where a victim row is sandwiched between two aggressor rows. The conventional \rh{} access pattern (\figref{fig:fast_hammer}(b)) consists of three steps. First, we issue an \gls{act} command to aggressor~1 (\circled{1}) and wait for \gls{tras}. Second, we issue a \gls{pre} command wait for $t_{RP}$ (\circled{2}) to ensure the DRAM bank is precharged. Third, we issue an \gls{act} command to aggressor~2 (\circled{3}) and wait for \gls{tras} before issuing a \gls{pre} command to precharge the bank (\circled{4}). These three steps comprise one \textit{hammer cycle}. Repeating the hammer cycle results in read disturbance bitflips in the victim row.
The \fastHammer{} access pattern (\figref{fig:fast_hammer}(c)) violates $t_{RP}$ and activates aggressor rows in quick succession. This results in directly reducing the time to complete one hammer cycle.

\opportunity{\fastHammer{} can enable fast RDT profiling by reducing the time to perform one hammer cycle.}

\subsection{RDT Reduction with \fastHammer{}}
\label{sec:rdt_reduction}
\begin{figure}[ht]
\centering
\includegraphics[width=0.87\linewidth]{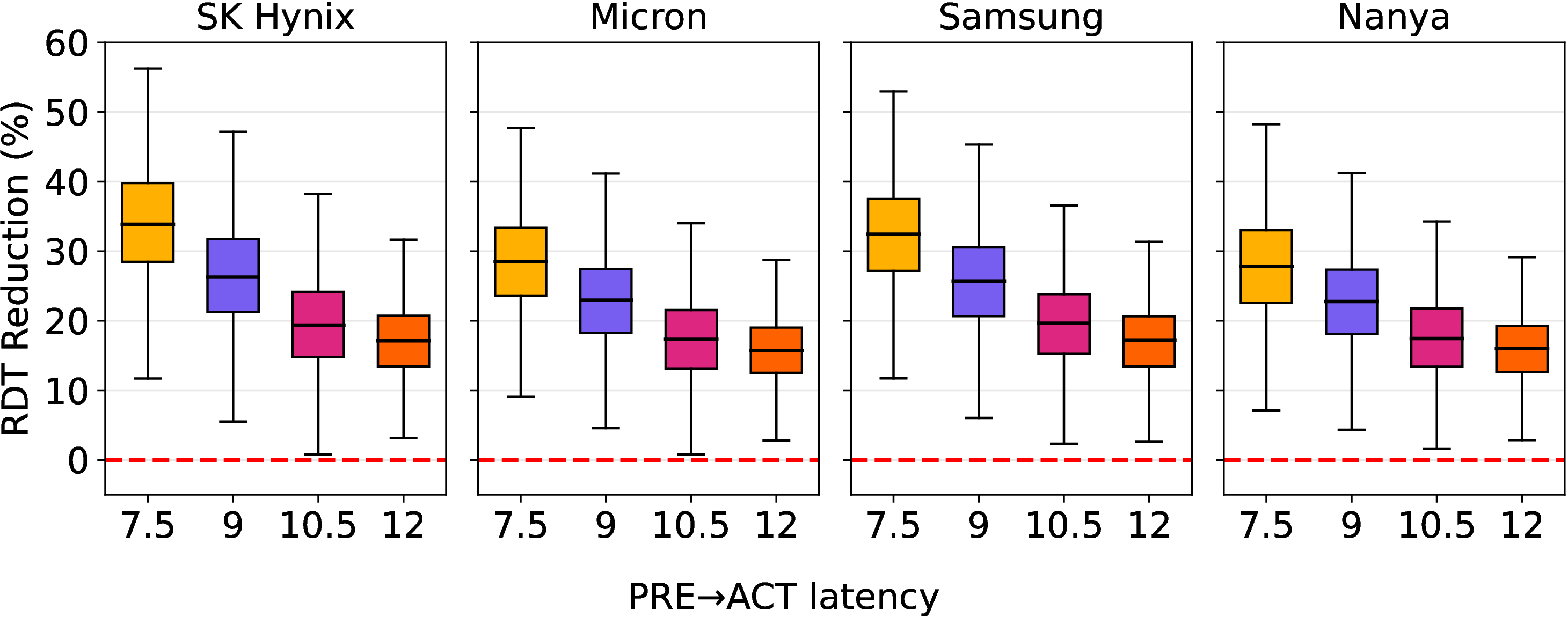}
\caption{RDT reduction with reduced PRE-to-ACT latency.}
\label{fig:rdt_reduction}
\end{figure}

\noindent{\textbf{RDT Reduction per DRAM Row.}}
We analyze the RDT of rows when performing double-sided RowHammer and \fastHammer{} with reduced \gls{pre}-to-\gls{act} latency. 
\figref{fig:rdt_reduction} shows the RDT reduction observed across \iey{all tested real DRAM chips} 
from four major manufacturers as subplots in a box-and-whiskers plot. 
The x-axis shows the \gls{pre}-to-\gls{act} latency in nanoseconds from 7.5 ns to 12 ns.\footnote{Prior works show that reducing the \gls{pre}-to-\gls{act} latency below 7.5 ns results in simultaneously activating multiple rows~\cite{yuksel2025pudhammer,yuksel2024simultaneous,yuksel2024functionally,gao2019computedram}. While this reduces the RDT~\cite{yuksel2025pudhammer}, the characteristics of the observed bitflips (e.g., the bitflip directionality) are different from RowHammer bitflip characteristics, indicating the leakage paths may differ. Therefore, we do \textit{not} reduce the \gls{pre}-to-\gls{act} latency below 7.5~ns to profile for RowHammer vulnerability.} The y-axis shows the RDT reduction in percentage compared to the RDT values of the rows with nominal \gls{pre}-to-\gls{act} latency, i.e., a conventional RowHammer double-sided access pattern. 

\observation{Reducing the \gls{pre}-to-\gls{act} latency reduces the per-row RDT, on average, across all modules from all manufacturers.}

\begin{figure*}[!t]
\centering
\includegraphics[width=0.86\linewidth]{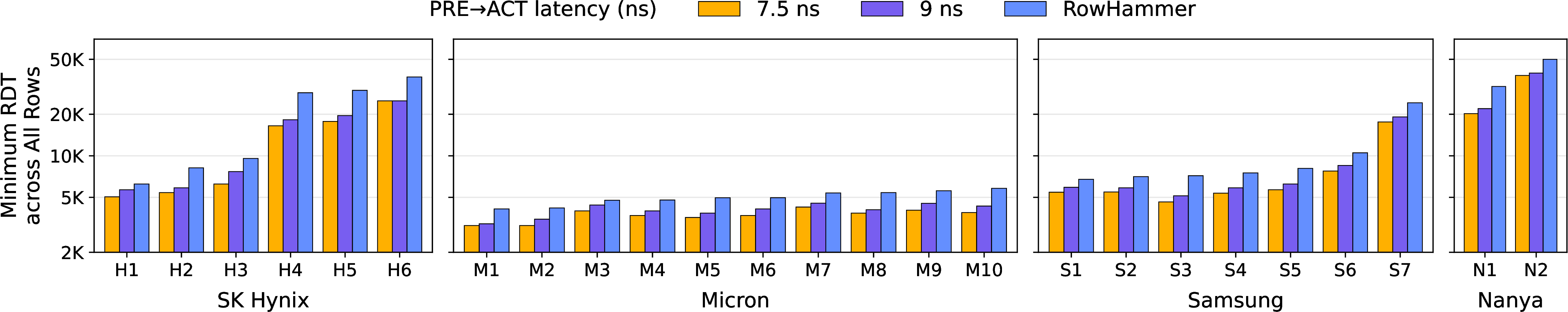}
\caption{Per module minimum RDT values with reduced PRE to ACT latency.}
\label{fig:min_rdt_reduction}
\end{figure*}

We observe that \fastHammer{} with \param{7.5~ns} \gls{pre}-to-\gls{act} latency reduces a row's RDT by \param{33.5\%}, \param{28.4\%}, \param{30.81\%}, and \param{27.0\%}, on average, for SK Hynix, Micron, Samsung and Nanya modules, respectively.
{We hypothesize that reducing the \gls{pre}-to-\gls{act} latency leaves more electrons emitted from traps near the just-closed aggressor wordline available when the opposite aggressor wordline turns on. The electric field generated by the opposite aggressor redirects more of these electrons into the victim cell’s storage node, increasing charge loss~\cite{yang2019trap,zhou2023double}. 3D TCAD simulations of double-sided \rh{} align with our observations: within the short-latency regime tested in our experiments, reducing the \gls{pre}-to-\gls{act} latency leaves more emitted electrons near the just-closed aggressor wordline during the opposite aggressor’s pulse-on phase, thereby reducing RDT~\cite{zhou2023double}.}

\observation{As the \gls{pre}-to-\gls{act} latency decreases, the RDT reduction increases significantly.} 

\nb{For example, we observe an average RDT reduction of \param{17.3\%}, \param{19.6\%}, \param{26.0\%} and \param{33.5\%} across the SK Hynix modules, at 12~ns, 10.5~ns, 9~ns and 7.5~ns, respectively. }

\opportunity{\fastHammer{} can enable fast RDT profiling by reducing \nb{the number of hammers to induce} a bitflip per row, on average.}

\noindent\textbf{\nb{RDT Reduction in the minimum RDT per DRAM Module.}}
\figref{fig:min_rdt_reduction} shows the per module minimum RDT values with reduced \gls{pre}-to-\gls{act} latency for four major manufacturers. The x-axis shows the \iey{all tested real} DRAM modules from each \iey{tested} manufacturer. The y-axis shows the minimum RDT across all rows within the \iey{tested} DRAM module \iey{on a} log scale. Each bar \iey{shows one PRE-to-ACT latency value}: the latency values with \iey{the} largest average \iey{RDT} reductions \nb{per row} (i.e., 7.5 ns and 9 ns) and the nominal latency.

\observation{The minimum RDT of each tested module \nb{decreases} with \nb{decreasing} \gls{pre}-to-\gls{act} latency.}

We observe that the minimum RDT observed in a module reduces by \param{21.7\%} and \param{28.3\%} at \param{9~ns} and \param{7.5~ns}, respectively, across all modules from all manufacturers. 

\opportunity{\fastHammer{} can identify a DRAM module's minimum RDT with a reduction.}

\subsection{Detailed Analysis of \fastHammer{}}
\label{sec:detailed}

To further understand the read disturbance behavior with \fastHammer{}, we extensively characterize \nb{representative} modules from four major manufacturers and analyze (i)~the RDT distribution, (ii)~the overlap of read disturbance-vulnerable cells, and (iii)~the \iey{maintaining} of the RDT ranking across rows. 

\begin{figure}[ht]
\centering
\includegraphics[width=0.87\linewidth]{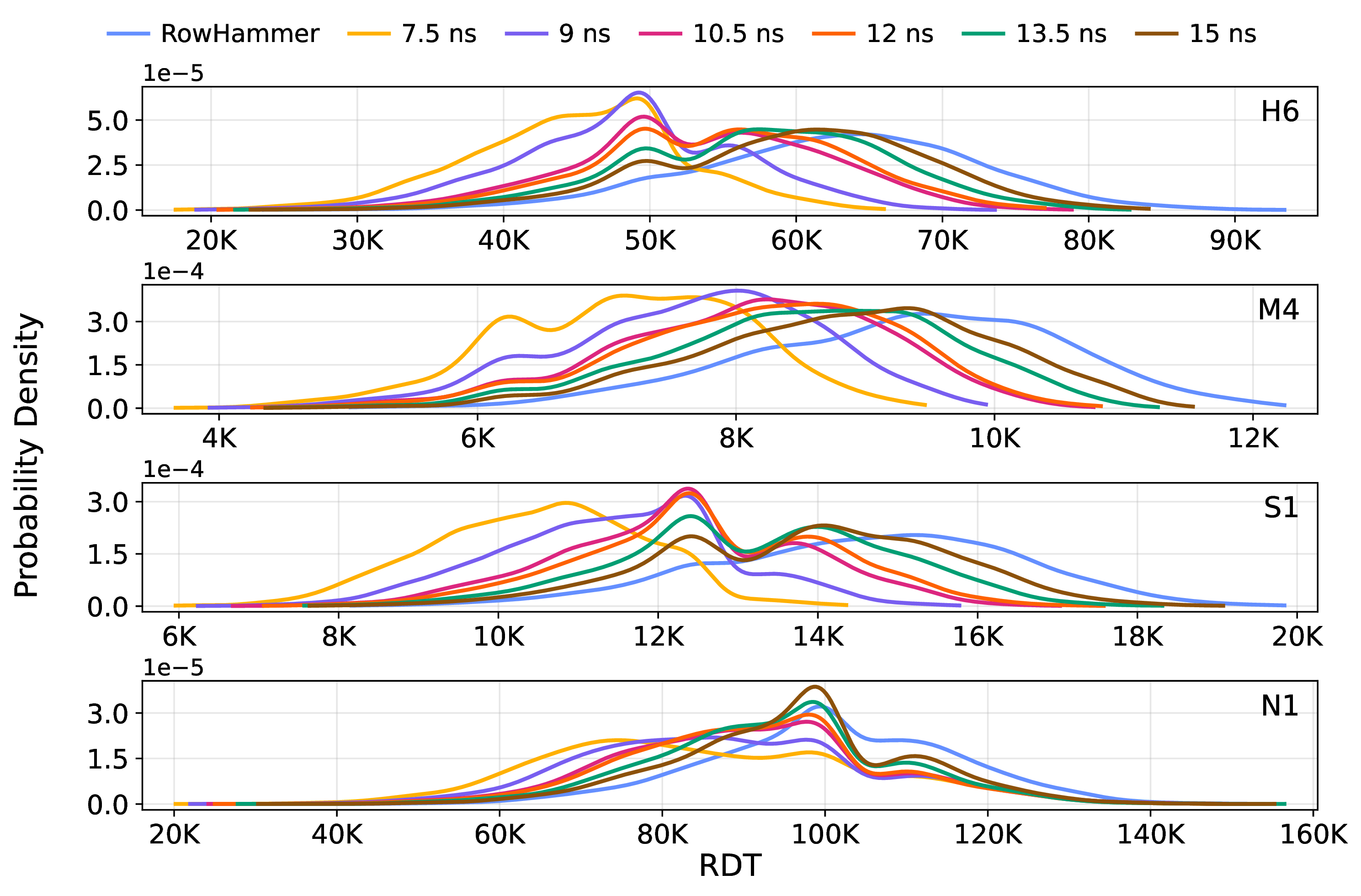}
\caption{Per module RDT distribution with reduced PRE to ACT latency.}
\label{fig:per_module_rdt_distribution}
\end{figure}

\noindent\textbf{RDT Distribution Shift with \fastHammer{}.}
\figref{fig:per_module_rdt_distribution} shows the distribution of RDTs \iey{across all tested rows} in four representative modules as subplots. The x-axis shows the RDT of an individual row, and the y-axis shows the probability density across the selected module's rows. Each line shows a \iey{different} \fastHammer{} latency from 7.5 ns to 15 ns and \rh{}. 

\observation{Reducing the \gls{pre}-to-\gls{act} latency shifts the RDT distribution of a module toward lower RDT values.}

The RDT distribution shifts toward lower RDT values as the \gls{pre}-to-\gls{act} latency decreases, in all four tested modules. The shift affects the entire distribution: both the lower tail, which determines a module's minimum RDT, and the bulk of the distribution move toward lower RDT values. At 7.5~ns, the distributions' \nb{median} values decrease by \param{28.2\%}, \param{23.6\%}, \param{28.0\%}, and \param{16.6\%} \nb{compared to  \rh{}} for the SK Hynix, Micron, Samsung, and Nanya modules, respectively.
\nb{We conclude that \fastHammer{} affects a module's RDT distribution by shifting it toward lower RDT values.}

\begin{figure}[ht]
\centering
\includegraphics[width=0.9\linewidth]{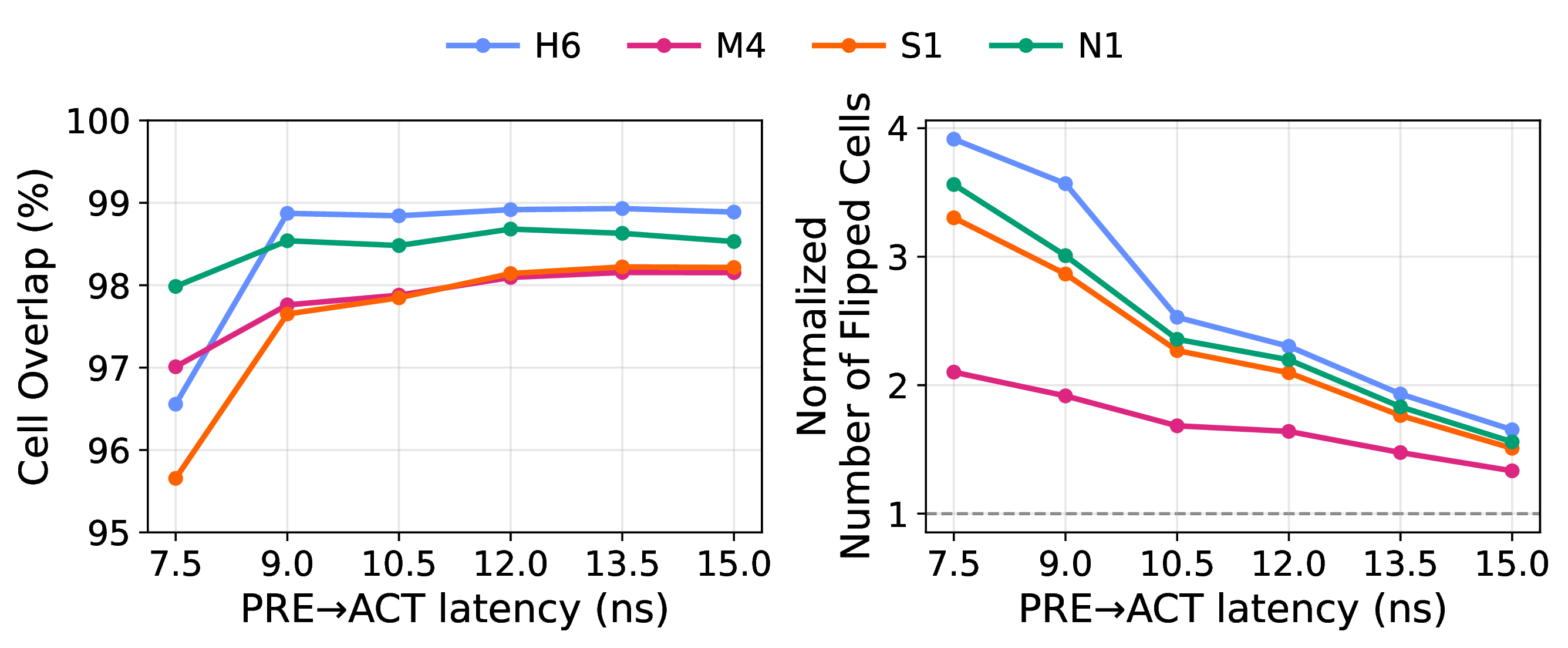}
\caption{Per module (a)~cell overlap (\%) with \rh{} and (b)~number of flipped cells normalized to \rh{}, with reduced PRE-to-ACT latency.}
\label{fig:per_module_cell_coverage}
\end{figure}

\noindent\textbf{Cell Overlap.}
To understand whether \fastHammer{} exacerbates \rh{}, 
\iey{we compare the set of cells that experience bitflips as we sweep the \gls{pre}-to-\gls{act} latency (i.e., for each \gls{pre}-to-\gls{act} latency, the set of cells that experience bitflips at that latency) with the set of cells that experience \rh{} bitflips (i.e., when the \gls{pre}-to-\gls{act} latency equals the nominal value). \figref{fig:per_module_cell_coverage} shows how reducing the \gls{pre}-to-\gls{act} latency (x-axis) changes (a)~the fraction of \rh{}-flipped cells that \fastHammer{} also flips (i.e., the cell overlap) and (b)~the number of cells that \fastHammer{} flips, normalized to the number of cells that \rh{} flips (y-axis) for four representative DRAM modules.}

\observation{\fastHammer{} flips nearly all cells that \rh{} flips, at every tested \gls{pre}-to-\gls{act} latency.}

\fastHammer{}'s cell overlap with \rh{} exceeds \param{95\%} at every tested latency in all tested modules. The overlap increases with the \gls{pre}-to-\gls{act} latency, from \param{95.7\%} at 7.5~ns to \param{98.9\%} at 15~ns, and remains nearly constant at $\geq$9~ns. 

\observation{\fastHammer{} flips more cells than \rh{}, and increasingly so as the \gls{pre}-to-\gls{act} latency decreases.}

\fastHammer{} flips up to \param{3.9}$\times$ more cells than \rh{} at 7.5~ns and \param{1.6}$\times$ more at 15~ns. Combined with the near-total cell overlap, this shows that \fastHammer{}'s bitflips form a superset of \rh{}'s bitflips: \fastHammer{} induces bitflips in the cells that \rh{} already flips and in additional cells. We hypothesize that this is due to the reduced \gls{pre}-to-\gls{act} latency exacerbating the \nb{read disturbance} leakage.

\noindent\textbf{Row Overlap at Top-N RDT Ranking.}
\label{sec:row_ranking}
\iey{We analyze whether \fastHammer{} can identify the rows that are most vulnerable to \rh{}, i.e., whether the \rh{}-weakest rows also appear among the \fastHammer{}-weakest rows. 
We rank all profiled rows in a module twice in ascending order of RDT (weakest row first): once by their \rh{} RDT and once by their \fastHammer{} RDT. We take the $N$ weakest rows of the \rh{} ranking (i.e., top-$N$) and measure the fraction of them that appear within the top $X\%$ of the \fastHammer{} ranking. We refer to this fraction as the \emph{row overlap}.}

\begin{figure}[ht]
\centering
\includegraphics[width=1\linewidth]{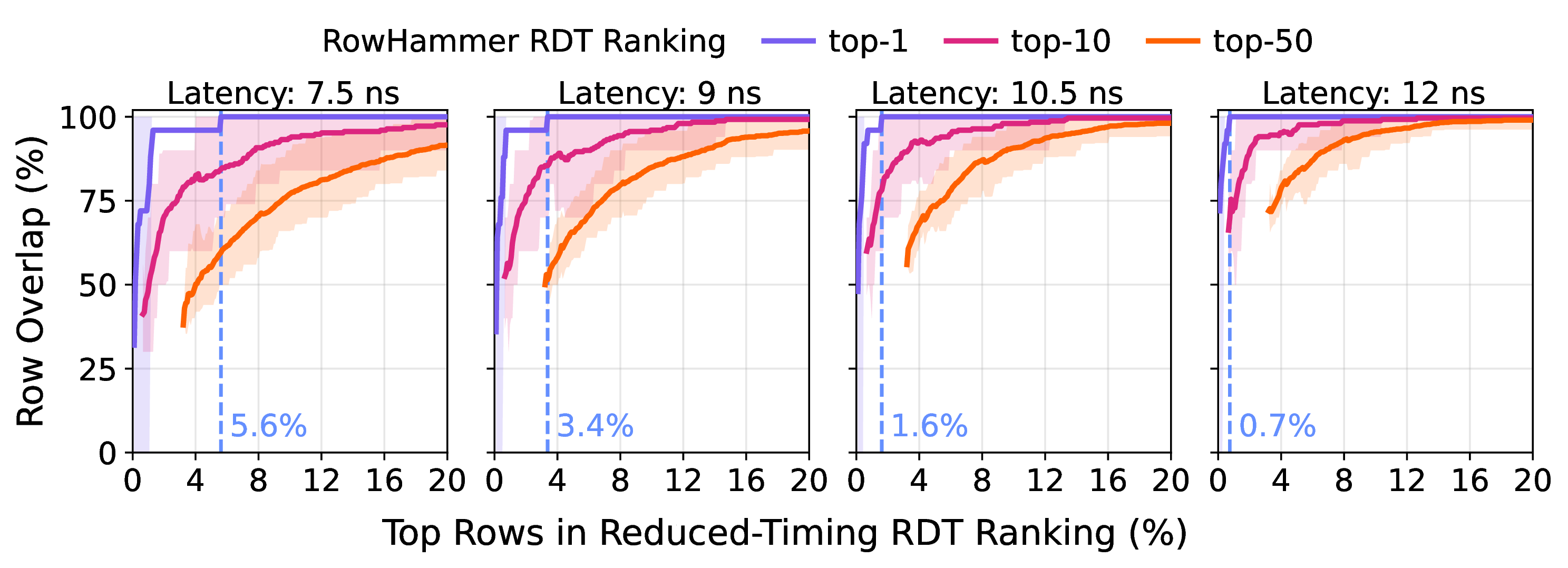}
\caption{\iey{Fraction of the $N$ weakest \rh{} rows (row overlap, y-axis) that fall
within the top $X\%$ of rows ranked by \fastHammer{} RDT (x-axis), for
four reduced \gls{pre}-to-\gls{act} latencies.}}
\label{fig:per_module_row_coverage}
\end{figure}

\figref{fig:per_module_row_coverage} shows the row overlap across \nMODULES{} real DRAM modules, with one subplot per \gls{pre}-to-\gls{act} latency. The x-axis is the fraction \iey{$X$} of the \fastHammer{} ranking we consider, as a percentage of all profiled rows;
the y-axis is the resulting row overlap. Each curve shows the mean overlap across all \nMODULES{} modules for one value of N (\param{1}, \param{10}, \param{50}), and the shaded region spans the \param{10th} to \param{90th} percentile across modules. For 
N=1, a module's overlap is either \param{0\%} or \param{100\%}, such that the purple curve gives the percentage of modules whose \rh{}-weakest row falls within the given fraction ($X\%$). The dashed vertical line marks the smallest fraction that captures the \rh{}-weakest row in \emph{every} tested module.

\observation{\fastHammer{}'s ranking places a module's \rh{}-weakest row among its highest-ranked rows.}

In \emph{all} tested modules, the \rh{}-weakest row falls within the top \param{5.6\%}, \param{3.4\%}, \param{1.6\%}, and \param{0.7\%} of the \fastHammer{} ranking at 7.5ns, 9ns, 10.5ns, and 12ns \gls{pre}-to-\gls{act} latency, respectively (dashed vertical lines).\label{resp:cq2_2}\labelcq{2}\revcommon{\footnote{\revcommon{To evaluate the effect of VRD on this ranking, we repeat the analysis in representative modules across \param{50} \fastHammer{} RDT measurements per row, and do \emph{not} observe a significant change in the row overlap (not shown).}}} Because the \rh{}-weakest row determines a module's minimum RDT, \fastHammer{}'s ranking \nb{contains} the row that sets this minimum \nb{within} a small, bounded region at the head of \nb{its} ranking.

\observation{\fastHammer{}'s row overlap degrades for larger sets of \rh{}-weak rows, and increasingly so as the \gls{pre}-to-\gls{act} latency decreases.}
\iey{At 12~ns, the top \param{20\%} of the \fastHammer{} ranking contains \param{99.0\%} of the top-\param{50} \rh{}-weakest rows, but only \param{91.5\%} at 7.5~ns.}
The same trend holds for the single weakest row: the region that contains it in every module widens by \param{8}$\times$, from \param{0.7\%} at 12~ns to \param{5.6\%} at 7.5~ns. This is consistent with our cell overlap analysis (\figref{fig:per_module_cell_coverage}): more aggressive \gls{pre}-to-\gls{act} latency reductions induce bitflips in an increasing number of cells beyond those \rh{} flips, which perturbs the relative ordering of rows by RDT.

\opportunity{\fastHammer{} preserves the RDT ranking of rows well enough to \nb{enable measuring the minimum \emph{\rh{}} RDT of a module (i.e., with \textit{no} RDT reductions due to reduced timing constraints) by reprofiling \textit{only} a small candidate set with \rh{}-based profiling}.}

\begin{figure}[ht]
\centering
\includegraphics[width=1\linewidth]{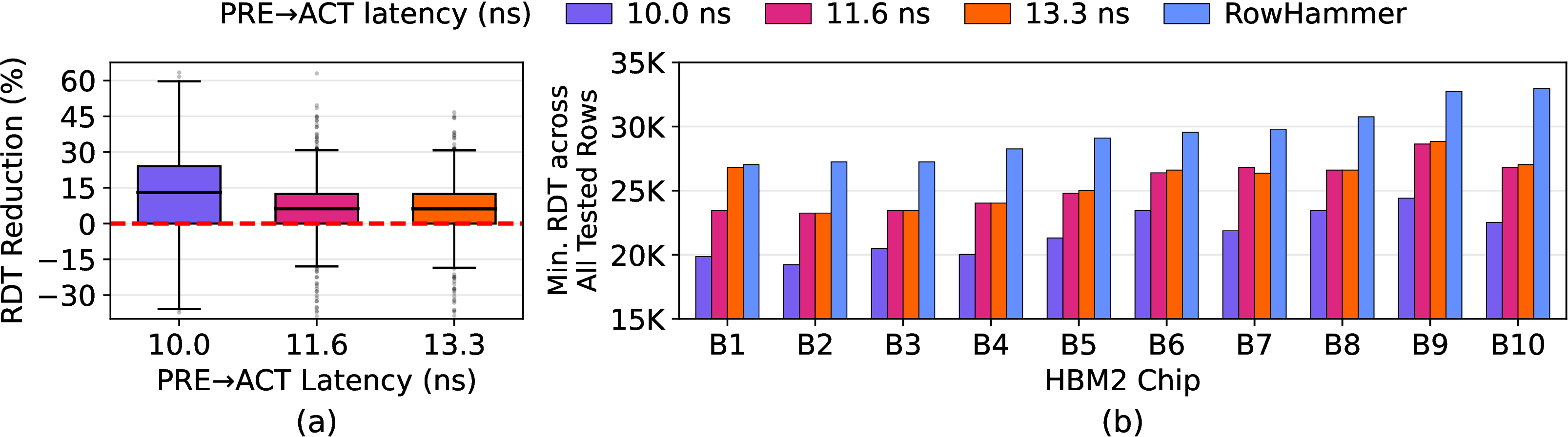}
\caption{\revd{(a) Per row RDT reduction and (b) per module minimum RDT reduction with \fastHammer{} in HBM2 DRAM chips.}}
\label{fig:hbm_analysis}
\end{figure}

\label{resp:d2}
\head{\revd{\fastHammer{} in HBM2}}\labeld{2} \revd{We experimentally characterize \fastHammer{}'s RDT reduction, and cell and row overlap characteristics in 10 HBM2 DRAM chips. \figref{fig:hbm_analysis} (a) shows the RDT reduction per DRAM row in all tested HBM2 chips and \figref{fig:hbm_analysis} (b) shows the minimum RDT values per HBM2 chip.}

\revd{We make three key observations. First, as \gls{pre}-to-\gls{act} latency decreases, the RDT reduction increases. We observe average RDT reductions of \param{12.6\%}, and \param{6.5\%} at 10.0 ns, and 13.3 ns, respectively. Second, the minimum RDT of each tested chip is lower with \fastHammer{} compared to the RowHammer access pattern (e.g., \param{26.5\%} average minimum RDT reduction at 10.0 ns).}
\revd{Third, we observe that at 10.0 ns, \fastHammer{} flips nearly all cells (\param{98.8}\%) that \rh{} flips and places a chip's \rh{}-weakest row among its top 0.1\% rows (not shown). We conclude that \fastHammer{} exhibits similar characteristics in HBM2 chips to DDR4 modules in terms of RDT reduction, and cell and row overlap.}
\section{\X{}: A Fast RDT Profiling Technique}
\label{sec:profiler}

We propose a new RDT profiling technique, \X{}. \X{}'s goal is to reduce the time required to profile a DRAM module's RDT. \X{} achieves this goal by hammering each row with reduced DRAM timing constraints (i.e., with \fastHammer{}) during profiling to exacerbate read disturbance intentionally and measure each row's RDT quickly. \X{} then uses these measurements to determine the module's minimum RDT, i.e., the value that existing \rh{} mitigations use to prevent bitflips. We describe two variants of \X{}:~(i)~\Xone{} reports a conservative estimate of the minimum RDT, and (ii)~\Xtwo{} improves the accuracy of the estimate at the cost of additional profiling time.

\head{High-Level Overview}
\label{sec:overview}
To profile a module's RDT, \X{} profiles each row and picks the lowest RDT measurement. For each row, \X{} performs a profiling routine that consists of three steps:
\textbf{(Step 1.)} \X{} initializes the victim and two aggressor rows with initial data patterns. 
\textbf{(Step 2.)} \X{} performs double-sided \fastHammer{} by repeatedly activating the aggressor rows to complete a predetermined hammer count.  
\textbf{(Step 3.)} \X{} reads the victim row's data to check for bitflips.

This routine is repeated multiple times to accurately identify the RDT in a binary search approach. \label{resp:e2}\labele{2}\reve{\X{} starts with an initial hammer count ($H_{init}$).} If the victim row does \emph{not} have a bitflip, \X{} increments the hammer count in Step 2 \reve{by an additive step that doubles after every routine (i.e., $H_{init}+\Delta$, $H_{init}+2\Delta$, $H_{init}+4\Delta$, \dots), up to the largest hammer count that fits within the refresh window}, and repeats the routine until a bitflip is observed. \reve{If $H_{init}$ already induces a bitflip, \X{} divides the hammer count by a factor $\beta$ until the victim row does \emph{not} have a bitflip. \X{} then performs a binary search between the highest hammer count without a bitflip and the lowest hammer count with a bitflip, testing their geometric mean in each step, until the two are within a relative tolerance $\epsilon$.} \X{} \reve{reports the lowest hammer count} that still results in a bitflip (i.e., \reve{the} \emph{minimum} hammer count to induce a bitflip). \reve{\X{} repeats this search $N_{search}$ times per row and determines the minimum result as the row's RDT.}
\reve{
Based on empirical data, we set $H_{init}$ to twice the module's minimum RDT measured in the previous profiling window (\param{10K} in the first window), $\Delta{=}\param{10K}$, $\beta{=}\param{1.2}$, $\epsilon{=}\param{1\%}$ (at least \param{10} hammers), and $N_{search}{=}\param{5}$.}

\head{\Xone{}} 
\Xone{} reports the minimum RDT observed by performing \fastHammer{} across the module. Doing so, \Xone{} profiles a DRAM module quickly by (i)~reducing the time it takes to perform one hammer and (ii)~reducing the number of hammers performed on average. As a result, \Xone{}'s reported RDT is a conservative estimate of the module's \rh{} RDT because \fastHammer{} reduces the RDT of a module (e.g., by \param{21.7\%}, on average, as shown in~\secref{sec:rdt_reduction}). 

\head{\Xtwo{}} 
\label{sec:hare-filter}
\label{resp:cq2}
To improve the RDT estimation of \fastHammer{}, \Xtwo{} leverages the row ranking property of \fastHammer{} (\secref{sec:detailed}):
{\fastHammer{} preserves the RDT ranking of rows such that a module's minimum \rh{} RDT can be identified by exhaustively profiling \emph{only} its top-ranked rows.}
\labelcq{2}\revcommon{In each profiling iteration,} \Xtwo{} filters the module's rows: it ranks a batch of rows ($N_{profile\_batch}$) by their \fastHammer{} RDT and selects the top rows with the lowest values ($N_{reprofile}$). \Xtwo{} then profiles \emph{only} the selected $N_{reprofile}$ rows with the conventional \rh{} access pattern and reports the minimum \rh{} RDT among them. The selected rows contain the module's \rh{}-weakest row because \fastHammer{} preserves the RDT ranking of rows. \Xtwo{} therefore reports the module's \rh{} RDT at the cost of profiling a small portion of the rows with slow conventional \rh{} access pattern. 
\revcommon{Because \Xtwo{} re-ranks all rows in every profiling iteration, it dynamically selects the weakest rows as their RDTs change over time.}

\head{Periodic RDT Profiling with \X{}}
\label{subsec:periodic_profiling}
We implement both variants of \X{} as a periodic profiling scheme to distribute the latency across the execution time and reduce latencies induced by profiling. \X{} profiles all rows of the module within a \textit{profiling window} ($\tprofw$), analogous to the refresh window ($\trefw$), and performs one profiling operation every \textit{profiling interval} ($\tprofi$), analogous to the refresh interval ($\trefi$):
\begin{equation}
\label{eq:profiling_interval}
\tprofi = \tprofw \bigg/ \left\lceil \frac{N_{total\_rows}}{N_{per\_operation\_rows}} \right\rceil
\end{equation}
where $N_{total\_rows}$ is the total number of rows to profile and $N_{per\_operation\_rows}$ is the number of rows \X{} profiles in one operation.

\head{Choosing a Configuration}
Based on our characterization of \nCHIPS{} real DDR4 DRAM chips, we select \fastHammer{} with 9~ns \gls{pre}-to-\gls{act} latency to quickly profile the RDT of DRAM rows because this configuration provides (i)~a high reduction in RDT and the total time required for one hammer cycle compared to the \rh{} access pattern, and (ii)~a high cell and row overlap percentages. We implement a version of \X{} that profiles one row in each bank during one profiling operation (i.e., every profiling interval). We sweep a set of the profiling window and interval values in~\secref{sec:security_analysis} and~\secref{sec:performance_analysis}. 

\section{Error Probability Analysis}
\label{sec:security_analysis}

\atb{\X{} improves the read disturbance tolerance of a memory system by enabling rapid identification of 
the minimum RDT of a DRAM module. For example, when the minimum RDT of a DRAM module 
reduces over time due to VRD~\cite{olgun2025variable}, the system can 1)~use \X{} to rapidly identify the new minimum 
RDT value, 2)~configure the read disturbance mitigation with the newly identified minimum RDT. 
Configuring the mitigation with the newly identified value would reduce the 
probability of a read disturbance bitflip, improving overall system reliability. To quantitatively
analyze the system reliability benefits of using \X{}, we develop a 
probabilistic error model for VRD-induced read disturbance bitflips based on our 
empirical read disturbance characterization results.}

\subsection{\atb{The Baseline} Error Model}
\label{subsec:baseline_error_model}

\noindent\textbf{\atb{RDT of a DRAM Row}.}
We draw each row's RDT from an empirical per-row RDT distribution~\cite{olgun2025variable}. Because a row's RDT changes over time, we additionally draw a per-row \emph{drift ratio} from an empirical VRD distribution and define the row's \emph{effective RDT} \atb{($\mathit{RDT}_{\mathit{eff}}$)} as its measured RDT divided by that ratio. \label{resp:a1}\labela{1}\reva{{We assume a row's effective RDT is always lower than the measured RDT to evaluate a conservative scenario where VRD-induced changes can result in a bitflip when a mitigation is configured with the measured RDT.}} As in \secref{sec:characterization}, we express a row's RDT in activations of its aggressor rows: a victim row exhibits a bitflip once its aggressor rows have been activated $k = \lceil \mathit{RDT}_{\mathit{eff}} \rceil$ times.

\noindent\textbf{Mitigation.} We model a generic deterministic counter-based \rh{} mitigation that maintains an activation counter per row and preventively refreshes a row's victim rows once its counter reaches a \emph{configured RDT} \atb{($\mathit{RDT}_{\mathit{conf}}$)}. We set the configured RDT to the minimum RDT measured across \atb{all rows in a} bank. The mitigation can employ a \emph{guardband} ($g$) to the measured minimum RDT across the bank as a countermeasure against VRD-induced RDT changes~\cite{olgun2025variable}:
\begin{equation}
\label{eq:configured_rdt}
\mathit{RDT}_{\mathit{conf}} = \min(\mathit{RDT}_{\mathit{measured}}) \times (1 - g)
\end{equation}

\noindent\textbf{Vulnerable DRAM Rows.} A mitigation is configured using \emph{measured} RDTs, whereas rows fail at their \emph{effective} RDTs. A victim row is \emph{vulnerable} if $k < \mathit{RDT}_{\mathit{conf}}$ (i.e., effective RDT is lower than the mitigation's configured RDT).

\noindent\textbf{A Read Disturbance Bitflip.}
The \atb{activation} counter \atb{of a DRAM row} resets once it reaches $\mathit{RDT}_{\mathit{conf}}$ \atb{(and the row's victims are refreshed)}, thus a vulnerable victim row exhibits bitflips after $k$, $k+P$, $k+2P$, \dots aggressor activations, where $P = \lceil \mathit{RDT}_{\mathit{conf}} \rceil$. \atb{We determine the number of bitflips exhibited by a vulnerable victim row by sampling an empirically-observed distribution. We distribute the exhibited bitflips uniformly at random across the DRAM row because we do \emph{not} observe a spatial correlation between these bitflips.}

\noindent\textbf{\atb{An} Uncorrectable Error.} We model a DRAM chip that employs on-die single-error-correcting ECC over \param{128}-bit data words~\cite{patel2020beer,patel2019understanding,kwon2017an,kwak2017an}. If two or more bitflips exhibited by a vulnerable row fall in the same 128-bit word, we consider that an uncorrectable error.

\subsection{\atb{Online Profiler}}

We model an RDT profiler as a periodic profiler that measures rows at the profiling interval ($\tprofi$, \secref{subsec:periodic_profiling}) and lowers the mitigation's configured RDT as it discovers rows with lower RDTs, recomputing $\mathit{RDT}_{\mathit{conf}}$ from the new minimum as in \eqref{eq:configured_rdt}. The first version of \X{}, \Xone{}, updates the configured RDT immediately: as soon as it profiles a row whose effective RDT is lower than the current configured RDT, it lowers the configured RDT to that row's value. The second version of \X{}, \Xtwo{}, does \emph{not} update the configured RDT immediately, because ranking rows requires a batch: it profiles $N_{profile\_batch}$ rows, re-profiles the top-ranked $N_{reprofile}$ rows, and only then lowers the configured RDT to the lowest value it measures. Therefore, the system retains a configured RDT value by the time it takes to complete the profiling of one batch at the least. In both variants the configured RDT only decreases, so each reduction permanently protects every row whose bitflip threshold is at or above the new value.
For \Xtwo{}, we select $N_{profile\_batch}$ and $N_{reprofile}$ based on our empirical analysis (\secref{sec:characterization}). 

\subsection{Monte Carlo Evaluation Methodology.}
\label{subsec:mc_methodology}

\atb{In our error model, an interplay of empirically measured, non-parametric distributions (per-row RDT, drift, and bitflips per event) and stateful elements (e.g., the read disturbance mitigation and online RDT profiler) determines the DRAM bank's uncorrectable error probability. Thus, we estimate the time to uncorrectable error via Monte Carlo simulation instead of analytical modeling.}

\noindent\textbf{Simulated Monte Carlo Trial.} We simulate one DRAM bank of \param{$65,536$} rows and express time in terms of DRAM row activations (e.g., \param{45}~ns per activation~\cite{jedec2017ddr4}). Each trial draws a fresh bank from the measured input distributions and executes a read-disturbance-intensive memory access pattern (described next) until the first UE or until $10^{14}$ row activations, whichever comes first. 
We assume \atb{that a DRAM} row\atb{'s RDT} is drifted \atb{at} the start of each \atb{Monte Carlo} trial. This is a conservative assumption because \reva{modeled} drift only lowers a row's RDT. 
\nb{We assume a bitflip in memory persists and there is no scrubbing.}

\noindent\textbf{Memory Access Pattern.} \atb{We consider a pessimistic access pattern for the modeled DRAM system whereby the access pattern repeatedly 1)~uniformly at random selects a victim row and 2)~activates the aggressor rows of the victim row $RDT_{conf}$ times.\footnote{\atb{Our goal is to evaluate a benign workload access pattern that 1)~has no knowledge of which victim rows are vulnerable and 2)~induces bitflips as quickly as possible by concentrating activations on one victim row's aggressors at a time. Adversarial access patterns (where an adversary has oracular knowledge of the most vulnerable victim rows at all times) could lead to a higher error probability across all evaluated system configurations.}}}
We implement the error model in \secref{subsec:baseline_error_model} as a Monte Carlo simulator and report population statistics over independently drawn DRAM banks. Table~\ref{tab:sim_config} lists our simulation parameters.

\noindent\textbf{Metrics.} We report the probability of an uncorrectable error within a time budget, $P(\mathit{UE} \le t)$, and \emph{mean time to an uncorrectable error} (MTTUE). 
A bank in which \atb{no} row is \atb{vulnerable} can \emph{never} experience a UE, \atb{therefore} we also report fraction of banks that can \atb{experience UEs}.

\noindent\textbf{Estimators.} A trial that reaches the \atb{trial time limit} has \emph{no} observed UE. We therefore compute MTTUE as the restricted mean survival time \atb{until the trial time limit}, which is exact under common \atb{trial time limit across trials} and equals the plain mean when \atb{all trials end before the time limit}.

\begin{scriptsize}
\begin{table}[ht]
\scriptsize
\centering
\caption{Monte Carlo simulation parameters.}
\label{tab:sim_config}
\renewcommand{\arraystretch}{0.6}
\begin{tabular}{ll}
\toprule
\textbf{Parameter} & \textbf{Value} \\
\midrule
DRAM bank size ($N_{total\_rows}$)    & \param{65'536} rows \\
Time per row activation               & tRC (e.g., \param{45}~ns~\cite{jedec2017ddr4}) \\
On-die ECC                            & \param{(136,128)} SEC, \param{128}-bit data words \\
\midrule
Guardband ($g$)                       & \param{0\%}--\param{50\%} RDT reduction~\cite{olgun2025variable} \\
\midrule
Profiling interval ($\tprofi$)        & \param{6.75}--\param{128}~ms \\
Rows per batch ($N_{profile\_batch}$) & \param{3'000} (\secref{sec:characterization}), \param{1'500}, \param{750} rows \\
Re-profiled rows ($N_{reprofile}$)    & \param{100} (\secref{sec:characterization}), \param{50}, \param{25} rows \\
\midrule
Trials per configuration              & \param{1'000'000} banks \\
\atb{Trial time limit}                & \param{$10^{14}$} activations \\
\bottomrule
\end{tabular}
\end{table}
\end{scriptsize}

\subsection{Uncorrectable Error Probability Analysis}

\begin{figure}[ht]
\centering
\includegraphics[width=0.77\linewidth]{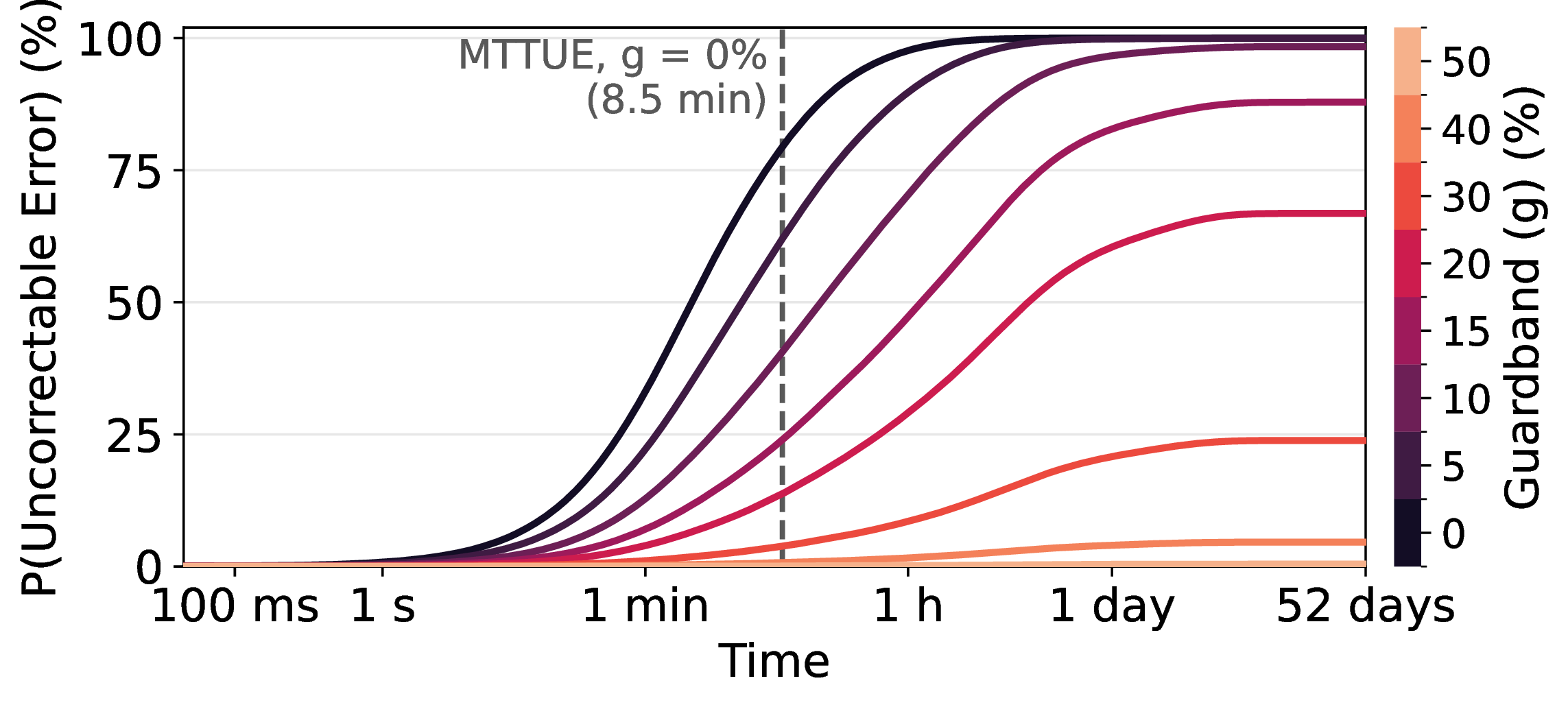}
\caption{P(Uncorrectable Error) percentage of a system with \textit{no} online profiler across time over 1M trials.}
\label{fig:p_ue_noprofiler}
\end{figure}

\noindent\textbf{No Online Profiling.}
\label{sec:pue_noprofiler}
\figref{fig:p_ue_noprofiler} shows the P(UE) percentage of a system that employs a \rh{} mitigation with \textit{no} online profiler. The x-axis shows time and the hue depicts different guardbands for the \rh{} mitigation. 

We make \param{two} key observations.
First, without a guardband ($g=0\%$, depicted as the black line), a system with \textit{no} online profiler experiences an P(UE) of $>$\param{99.9\%}, with
a mean time to uncorrectable error (MTTUE) (shown with the dashed gray line) of only \param{8.5 minutes}.
Second, a larger guardband reduces the P(UE), but \textit{only}
a very large guardband reduces it close to \textit{zero}. P(UE) reaches at most  \param{88.0\%}, \param{66.8\%},
\param{24.0\%}, \param{4.0\%}, and \param{0.5\%} for $g$=$15\%$, $20\%$,
$30\%$, $40\%$, and $50\%$, respectively. This is because a row experiences
bitflips \textit{only} if its RDT decreases by more than the guardband due to VRD, and
these bitflips become uncorrectable \textit{only} if ECC \textit{cannot} correct them. A larger
guardband means that fewer rows have an RDT decrease larger than the guardband,
thus, fewer trials end in an UE.

We conclude that a \rh{} mitigation that uses a static RDT based on one-time
profiling \atb{has a very high probability of encountering an UE.} 
Reaching a negligible P(UE) requires $g=50\%$, i.e., halving the RDT that the mitigation is
configured with, at a significant performance cost~\cite{olgun2025variable}.

\begin{figure}[ht]
\centering
\includegraphics[width=1\linewidth]{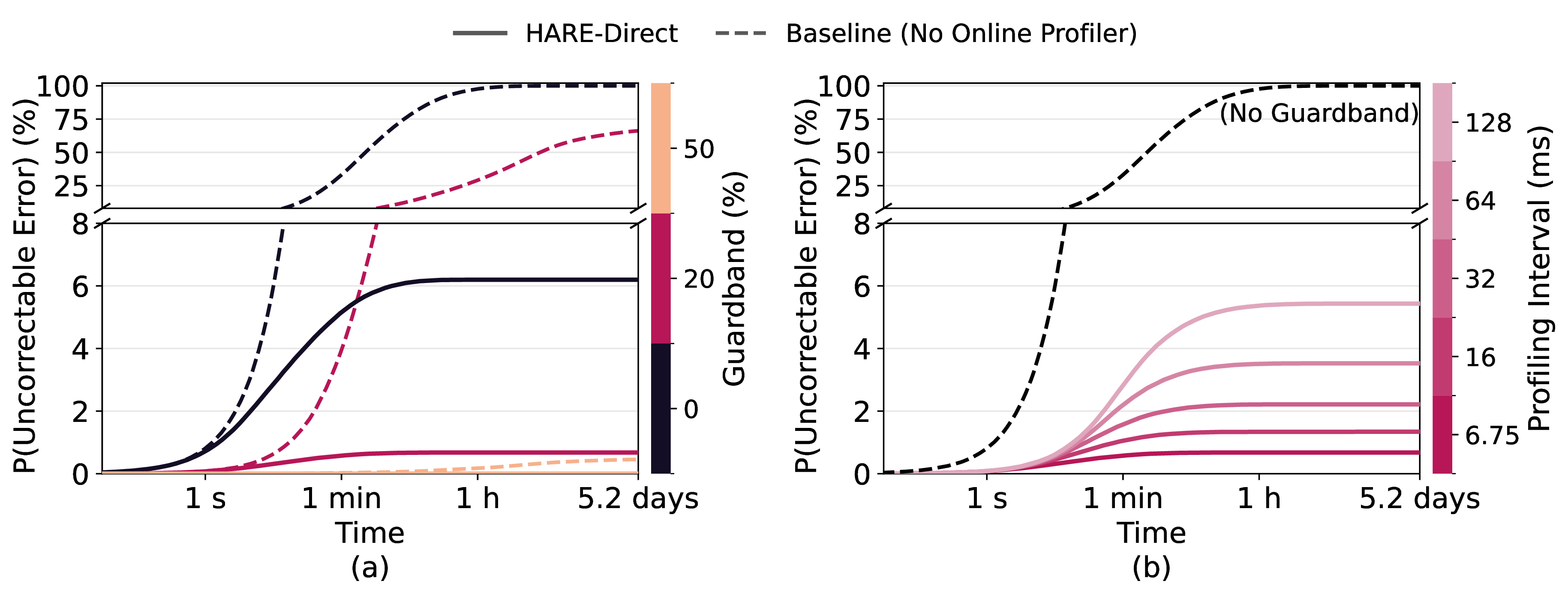} 
\caption{P(Uncorrectable Error) percentage of a system with \Xone{} and with
\textit{no} online profiler across (a) different guardband values at a profiling
interval of \param{6.75~ms}, and (b) different profiling intervals, over 1M trials.}
\label{fig:p_ue_profiler}
\end{figure}

\noindent\textbf{\Xone{}.}
\figref{fig:p_ue_profiler} shows the P(UE) of a system with
\Xone{} (solid) and with \textit{no} online profiler
(dashed) across time (cut at 5.2 days where curves are relatively horizontal). \figref{fig:p_ue_profiler}(a) depicts the P(UE) in systems employing \rh{} mitigations with guardband values of \param{0\%, 20\%, and 50\%}. The online profiler has a profiling interval of \param{6.75~ms}, the interval needed to profile all rows
within the MTTUE of the system with no guardband.

We make
\param{two} key observations.
First, online profiling reduces the P(UE) by more than
an order of magnitude at every guardband value. At $g$=$0\%$,
\Xone{} experiences a UE in \param{6.2\%} of the trials,
whereas the system with \textit{no} online profiler experiences a UE in \param{99.9\%} of the
trials. 
Second, online profiling reaches a low P(UE) with a
small guardband. \Xone{} experiences a UE in \param{0.67\%} of
the trials at $g$=$20\%$. Further increasing the guardband to \param{50\%} reduces \Xone{}'s P(UE) to \param{0.002\%}.
We conclude that \Xone{} significantly reduces the P(UE) compared to a system with \emph{no} online profiler, increasing with guardband. 

\figref{fig:p_ue_profiler}(b) shows the P(UE) of a system with \Xone{} profiling DRAM rows at various profiling intervals (with $g$=$20\%$, similar to the average RDT reduction observed in our empirical results~\secref{sec:characterization}) and a system \textit{without} an online profiler (and \textit{without} a guardband). We make two key observations. First, with increased profiling interval of 128~ms, \Xone{} still provides smaller P(UE) compared to a system with \textit{no} online profiler. Second, the P(UE) increases as the profiling interval increases. This is because, as profiling interval increases, \Xone{} discovers the minimum effective RDT later in time, leaving more time for the access pattern to trigger UEs.
We conclude that, when configured with a larger profiling interval, \Xone{} still provides a smaller P(UE) over a system with no online RDT profiler.

\begin{figure}[ht]
\centering
\includegraphics[width=0.77\linewidth]{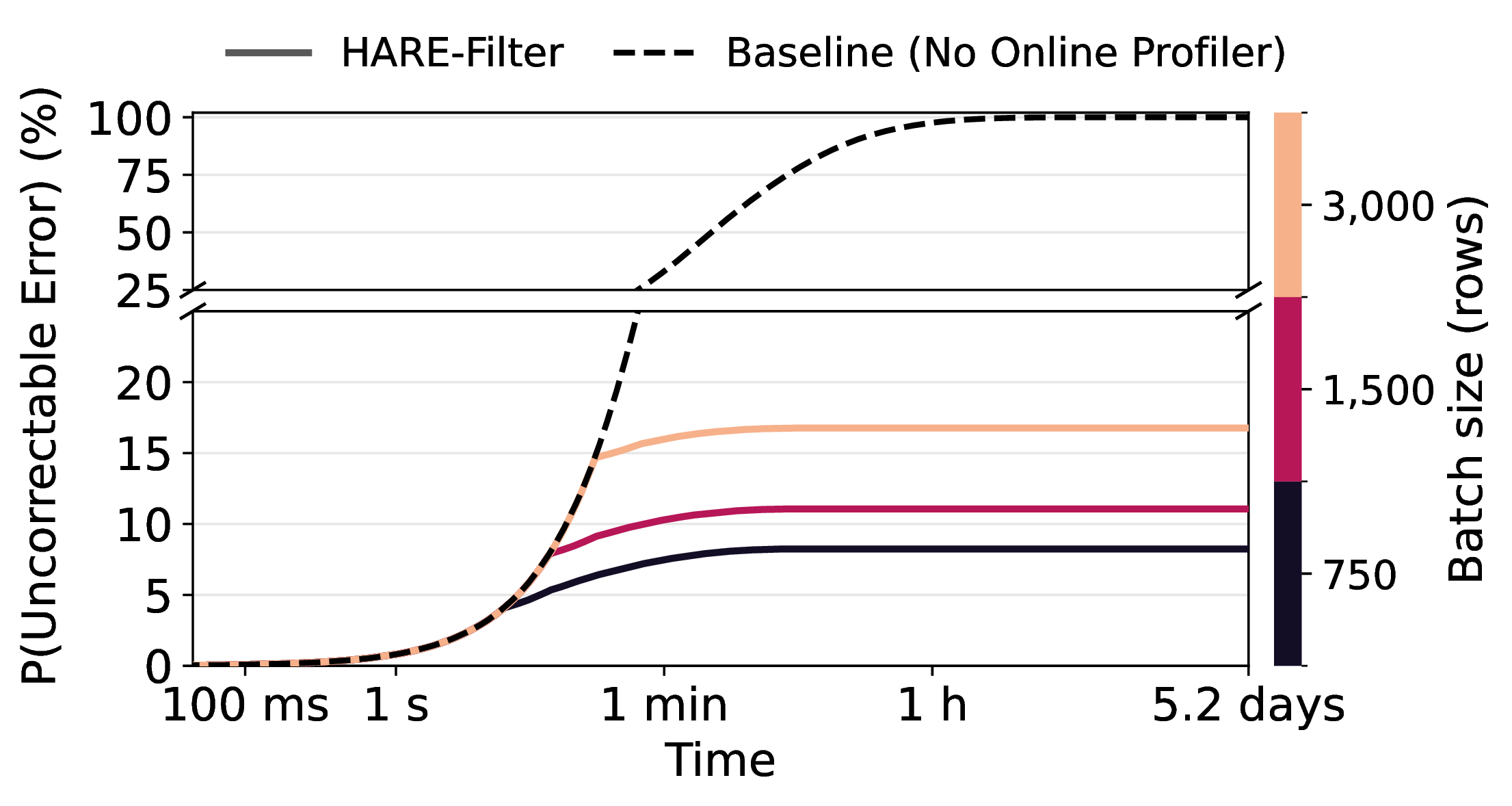}
\caption{P(Uncorrectable Error) percentage of a system with \Xtwo{} and
\textit{no} online profiler across \param{profiling batch size}, over 1M trials.}
\label{fig:p_ue_profiler_filter}
\end{figure}

\noindent\textbf{\Xtwo{}.}
We evaluate the reliability of a system employing \Xtwo{}. \Xtwo{} updates the \rh{} mitigation's configured RDT only after (i)~profiling a batch of rows ($N_{profile\_batch}$) with \fastHammer{} and (ii)~reprofiling a small percentage of the batch of rows ($N_{reprofile}$) with \rh{}. \figref{fig:p_ue_profiler_filter} shows the P(UE) of a system with \Xtwo{} employing various batch sizes, and with \emph{no} online profiler. We sweep $N_{profile\_batch}$ from 3K rows to 750 rows. We make \param{three} key observations. 
First, with the tested batch sizes, \Xtwo{} provides lower (by \param{12.1x} with 750-row batch) P(UE) compared to the system with \textit{no} online profiler. Second, \Xtwo{}'s P(UE) monotonically decreases as $N_{profile\_batch}$ decreases. This is because a smaller batch shortens the window before the first configuration update and most of the failing banks fail before any update has been applied. Third, \Xtwo{} can more accurately measure the \rh{} RDT of a row compared to \Xone{} (\secref{sec:profiler}) but it also results in an P(UE) larger than that of \Xone{} (not shown in the figure) due to the slower RDT updates. For example, at $N_{profile\_batch}$=750, \Xtwo{} has a 33.3\% higher P(UE) compared to \Xone{}.
We conclude that \Xtwo{} with the tested batch sizes improves the P(UE) over a system with \emph{no} profiler, while trading off RDT measurement accuracy with higher P(UE).

\section{Performance and Energy Consumption\\Analysis of \X{}}
\label{sec:performance_evaluation}

\begin{figure*}[!h]
\centering
\includegraphics[width=0.87\textwidth]{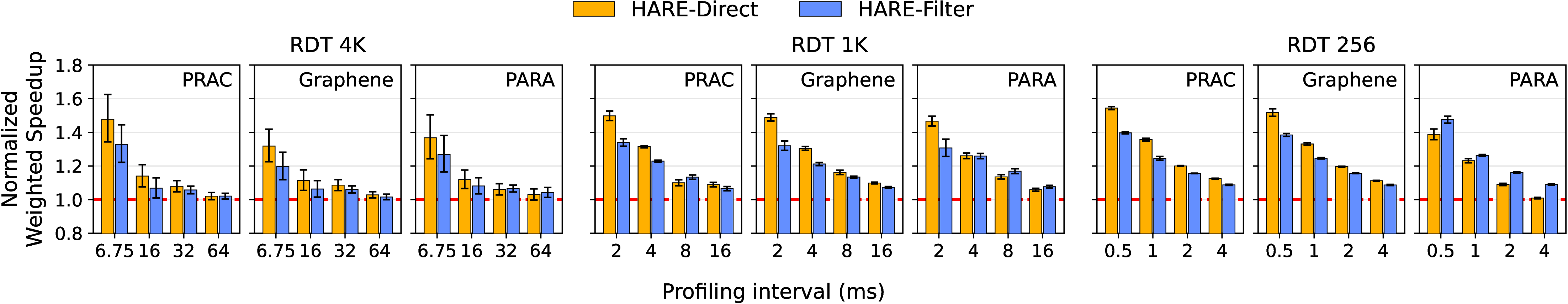}
\caption{Performance of a system using \Xone{} and \Xtwo{} normalized to the \rh{} profiler, across profiling intervals for three RDT values (groups of subplots) and three underlying \rh{} mitigations (subplots within each group).}
\label{fig:perf_all}
\end{figure*}

\subsection{System Performance Evaluation}
\label{sec:performance_analysis}
We evaluate \X{}'s impact on performance using Ramulator 2.0~\cite{luo2023ramulator2,ramulator2github}, a cycle-level DRAM simulator.
Table~\ref{table:system_configuration} shows the configuration of our simulated system.
We evaluate \X{} using 10 quad-core workload mixes.
To synthesize each workload mix, we randomly select workloads from the SPEC CPU2006~\cite{spec2006}, SPEC CPU2017~\cite{spec2017}, TPC~\cite{tpcweb}, MediaBench~\cite{fritts2009media}, and YCSB~\cite{ycsb} suites, considering only memory-intensive workloads with more than 10 row-buffer misses per kilo-instruction: \nb{519.lbm, 459.GemsFDTD, 450.soplex, h264\_decode, 520.omnetpp, 433.milc, 434.zeusmp, bfs\_dblp, 429.mcf, 549.fotonik3d, 470.lbm, bfs\_ny, bfs\_cm2003, 437.leslie3d, gups}.
\revd{We design the conventional \rh{} profiler,\labeld{4}\label{resp:d4} based on the RDT measurement
methodology widely used by prior characterization
works~\cite{orosa2021deeper,yaglikci2022understanding,luo2023rowpress,olgun2025variable}
(\secref{sec:methodology}). This is because,
to our knowledge, no prior work proposes an online profiler for \rh{}.
The \rh{} profiler employs the same periodic profiling scheme \atb{as HARE-Direct} and replaces the \fastHammer{} with \rh{} access pattern. Since \rh{} access pattern directly measures the \rh{} RDT, \rh{} profiler does \textit{not} require a second reprofiling phase and can directly update the RDT configuration of the underlying \rh{} mitigation.}

\newcolumntype{C}[1]{>{\let\newline\\\arraybackslash\hspace{0pt}}m{#1}}
\begin{table}[ht]
\scriptsize
\centering
\caption{{Simulated} System Configuration}
\begin{tabular}{l|C{5.8cm}}
 \hline
 \textbf{Processor} & {\SI{4.2}{\giga\hertz}, 4-core, 4-wide issue, {128-entry} instr. window} \\ \hline
 \textbf{Last-Level Cache} & {64-byte} cache line, 8-way {set-associative, \SI{8}{\mega\byte}} \\ \hline
 \textbf{Memory Controller} & {64-entry read/write request queues; Scheduling policy: FR-FCFS+Cap of \param{4}~\cite{mutlu2007stall}}; Address mapping: MOP~\cite{kaseridis2011minimalistic} \\ \hline
 \textbf{Main Memory} & DDR5 DRAM~\cite{ramulator2github}, 1 channel, 2 ranks, 8 bank groups, 4 banks/bank group, {64K} rows/bank\\ \hline
 \end{tabular}
 \label{table:system_configuration}
\end{table}

\noindent\textbf{Performance Impact of \X{}.}
\nb{We evaluate a system that employs a state-of-the-art \rh{} mitigation mechanism (i.e., PRAC~\cite{jedec2024jesd795c,canpolat2024understanding}, Graphene~\cite{park2020graphene}, and PARA~\cite{kim2014flipping}) with (i)~\Xone{}, (ii)~\Xtwo{} and (iii)~a \rh{} profiler.}
\nb{We configure all \rh{} mitigations with their corresponding preventive action thresholds, calculated based on the RDT configuration evaluated. We evaluate \Xone{} with a conservative RDT value with the average reduction of \param{21.7\%}, as observed in our empirical results (\secref{sec:characterization}).}
\figref{fig:perf_all} presents the performance of a system using \Xone{} and \Xtwo{} across different values of RDT, underlying \rh{} mitigations, and profiling intervals.
The plots of \figref{fig:perf_all} are organized into groups, with each group corresponding to a different RDT value (from left to right: 4K, 1K, and 256).
Each plot within a group corresponds to a different underlying \rh{} mitigation mechanism (from left to right: PRAC~\cite{jedec2024jesd795c,canpolat2024understanding}, Graphene~\cite{park2020graphene}, and PARA~\cite{kim2014flipping}).
The x-axis shows the profiling interval ($\tprofi$, i.e., time between consecutive profiling operations), while the y-axis shows weighted speedup normalized to the same system using a \rh{} profiler 
at the same $\tprofi$.
\nb{For RDT=4K, we sweep $\tprofi$ values of 6.75~ms (based on our analysis in~\secref{sec:security_analysis}), 16, 32, and 64. For lower RDT values, we scale the $\tprofi$ accordingly to simulate similar system reliability benefits for the online profilers.}

We make \param{four} key observations.
First, \Xone{} and \Xtwo{} outperform the \rh{} profiler, on average across all applications, for \emph{all} \nb{\rh{} mitigation, $\tprofi$, and RDT configurations.}
\Xone{}'s and \Xtwo{}'s average performance improvements across all
applications vary from \param{54.4\%} to \param{0.9\%} and \param{47.5\%} to \param{1.5\%}, respectively.
This is because both \Xone{} and \Xtwo{} profile rows faster than the \rh{} profiler, reducing the time spent on profiling and increasing the time available for demand memory requests.
Second, the speedup of both variants increases with decreasing $\tprofi$, \nb{as profiling consumes a larger fraction of total execution time, such that reducing profiling latency yields proportionally larger performance gains.} 
For example, with PRAC at RDT 256, \Xone{}'s (\Xtwo{}'s) speedup \nb{increases} from \param{12.5\%} (\param{8.8\%}) at $\tprofi{}{=}4$~ms to \param{54.4\%} (\param{39.7\%}) at $\tprofi{}{=}0.5$~ms.
Third, \Xone{} outperforms \Xtwo{} in most (\param{26} out of \param{36}) configurations, as its faster profiling operations outweigh the extra preventive actions caused by its conservative RDT estimate.
\nb{Fourth, \Xtwo{} outperforms \Xone{} in configurations where \rh{} mitigation overhead impacts system performance more than the profiling overhead.} For example, in a system that employs PARA, \Xtwo{} outperforms \Xone{} at every profiling interval, at the low RDT of 256. 
This is due to PARA's significant performance overheads for low \rh{} thresholds, which amplify the impact of \Xone{}'s conservative RDT estimate and resulting unnecessary preventive actions. 
\nb{The remaining configurations are those with long $\tprofi$ values at
\param{RDT${\leq}$1K}, where profiling operations are infrequent and thus, \rh{} mitigation overhead outweighs profiling cost.}

We conclude that \X{} improve system
performance over the conventional \rh{} profiler, and that its benefits are largest when profiling is frequent, i.e.,
at the short profiling intervals. \Xone{} performs better when profiling
overhead dominates, whereas \Xtwo{} is preferable when the mitigation's
overhead dominates, as its more accurate RDT estimate avoids unnecessary
preventive actions.

\begin{figure}[!h]
\centering
\includegraphics[width=\linewidth]{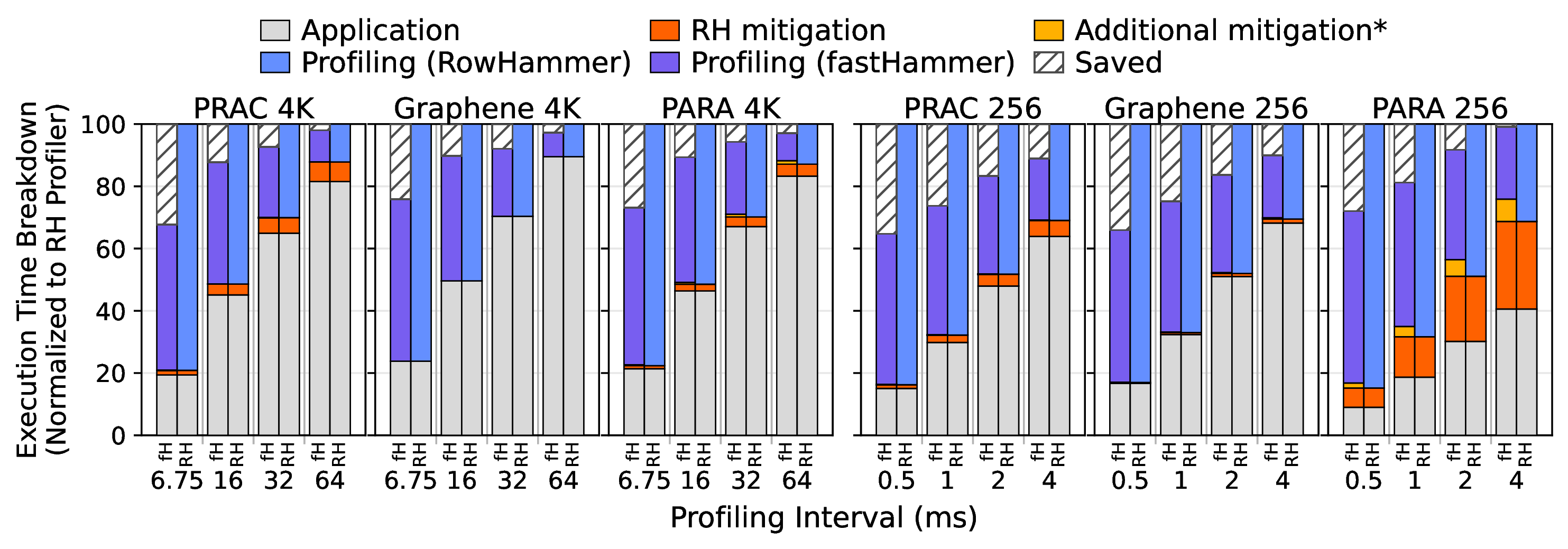}
\caption{\reve{Execution time breakdown of a system using \Xone{}, normalized to \rh{} profiler at RDT=4K and 256 across profiling intervals.}}
\label{fig:perf_breakdown}
\end{figure}

\head{\reve{Execution Time Breakdown Analysis}}\labele{1}\label{resp:e1}
\reve{\figref{fig:perf_breakdown} shows the execution time breakdown of a system that uses \Xone{} (fH) or the \rh{} profiler (RH), normalized to the system with the \rh{} profiler at RDT=4K and RDT=256 (subplots), across profiling intervals (x-axis). Each bar breaks execution time into application execution, \rh{} mitigation overhead, and profiling overhead. The \Xone{} bar also shows 1)~the \emph{additional mitigation} overhead caused by configuring the mitigation with \fastHammer{}'s lower RDT, and 2)~the execution time \Xone{} \emph{saves} compared to the \rh{} profiler (hatched).}

\reve{We make two key observations. First, \Xone{}'s speedup comes from reduced profiling time, which is largest at short profiling intervals (e.g., \Xone{} reduces execution time by \param{35\%} with PRAC at RDT=256 and $\tprofi{}{=}0.5$~ms). Second, \Xone{}'s additional mitigation overhead is negligible ($<$\param{1\%}) for PRAC and Graphene, and noticeable \emph{only} for PARA. We conclude that \Xone{}'s profiling time savings outweigh its additional mitigation overhead in \emph{every} evaluated configuration.}

\begin{figure}[ht]
\centering
\includegraphics[width=0.9\linewidth]{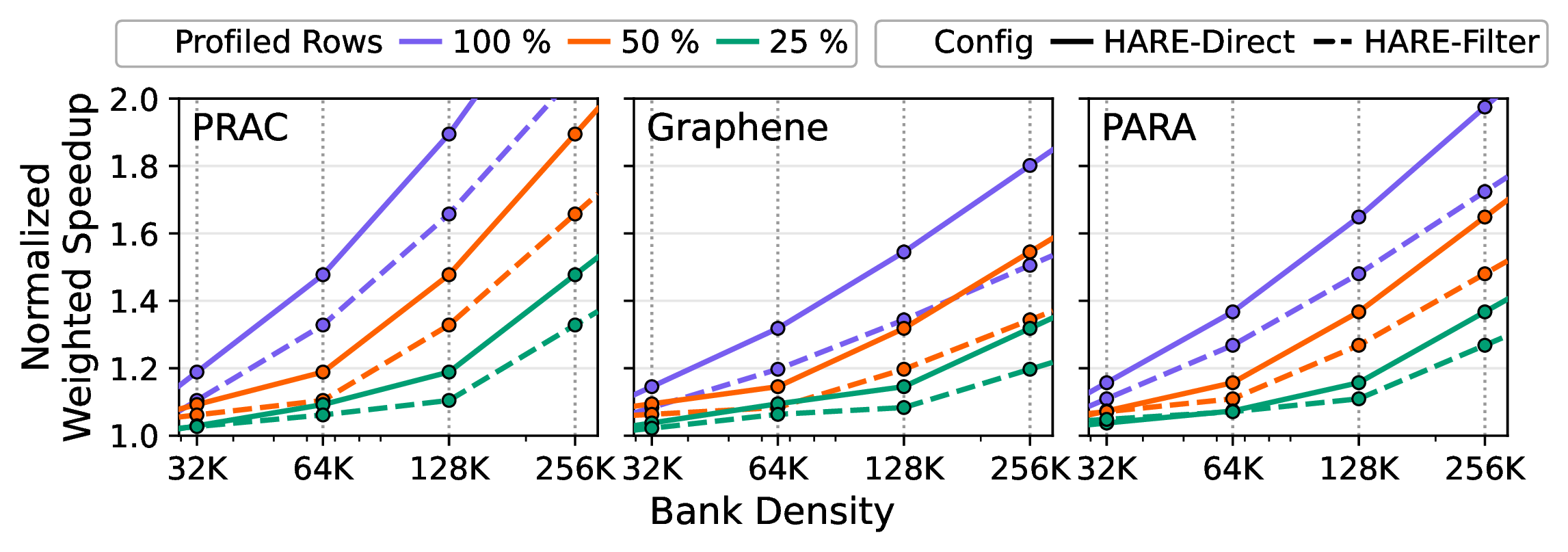}
\caption{\reva{Performance of \X{} normalized to the \rh{} profiler across bank densities, when profiling \param{100\%}, \param{50\%}, and \param{25\%} of the rows per bank.}}
\label{fig:bank_density}
\end{figure}

\head{\reva{Sensitivity to Bank Density and Profiled Rows}}
\labela{2}
\label{resp:a2_2}
\reva{DRAM bank density increases with chip density (e.g., from \param{64K} to \param{256K} rows per bank in \param{16Gb} to \param{64Gb} DDR5 chips~\cite{jedec2024jesd795c,kim2026pvac}), increasing the number of rows a profiler must profile. We evaluate \X{} across bank densities of \param{32K} to \param{256K} rows. To analyze the sensitivity to the number of profiled rows, we evaluate a hypothetical scenario where profilers profile \param{50\%} or \param{25\%} of the rows in each bank. {This scenario is optimistic as our results (\secref{sec:motive-spatial}, \figref{fig:row-sweep}) show that a module's minimum RDT does \emph{not} stabilize within \param{9K} profiled rows, so profiling a subset of rows may \emph{not} guarantee finding the minimum RDT.} \figref{fig:bank_density} shows the weighted speedup of \Xone{} (solid) and \Xtwo{} (dashed) normalized to the \rh{} profiler that profiles the same rows, at RDT=4K.}

\reva{ We make two key observations. First, \Xone{} outperforms the \rh{} profiler in \emph{every} configuration, and its benefit grows with bank density (e.g., with PRAC, from \param{47.7\%} at \param{64K} rows to \param{89.5\%} at \param{128K} rows).
Second, even in the optimistic scenario, profiling a subset of rows does \emph{not} eliminate the profiling cost in dense banks, e.g., even when profiling only \param{25\%}
of the rows, \Xone{} improves performance by \param{48\%} in a
\param{256K}-row bank.  We conclude that \X{} significantly reduces RDT profiling cost, and that this benefit grows as banks become denser.}

\subsection{Hardware Cost Analysis}

\X{} can be implemented in various ways, and we implement a version in the memory controller.
\Xone{} stores the current minimum RDT in a single register, thus, incurring negligible overheads.
\Xtwo{} maintains a per-bank \emph{lowest-RDT table} of the \param{128} rows with the lowest \fastHammer{}-measured RDT, each entry holding a \param{20}-bit RDT (up to ${\approx}$\param{1M} activations) and a \param{17}-bit row ID.
For the system of Table~\ref{table:system_configuration}, the storage overheads of \Xtwo{} are \param{37.2}~KiB.
We implement the lowest-RDT table in Verilog HDL and use open-source tools~\cite{ajayi2019toward,nangate45} to synthesize it and evaluate its latency, and area overheads.
\Xtwo{} incurs \param{0.17}~mm$^2$ area overhead.
Updating the lowest-RDT table takes \param{260}~ns (in the worst case).
Table updates are performed off-the-critical-path of the memory controller, introducing no additional latency to memory requests.
Since the shortest profiling interval we evaluate is $\tprofi{}{=}\param{0.5}$~ms, the table can always insert a new measurement before the next profiling operation.

\subsection{Energy Analysis}
\label{sec:energy_analysis}

\noindent\textbf{Methodology.} We measure the power consumption of a representative real DDR4 module at its supply rails while the module continuously executes each access pattern. We monitor both the core ($V_{DD}$) and the wordline  ($V_{PP}$) voltage/current rails, and compute the module's power. We compute the energy consumption of each profiling operation as the measured mean power multiplied by {the time required to complete a hammer cycle} and the hammer count to induce the first bitflip. 

\begin{figure}[ht]
\centering
\includegraphics[width=0.8\linewidth]{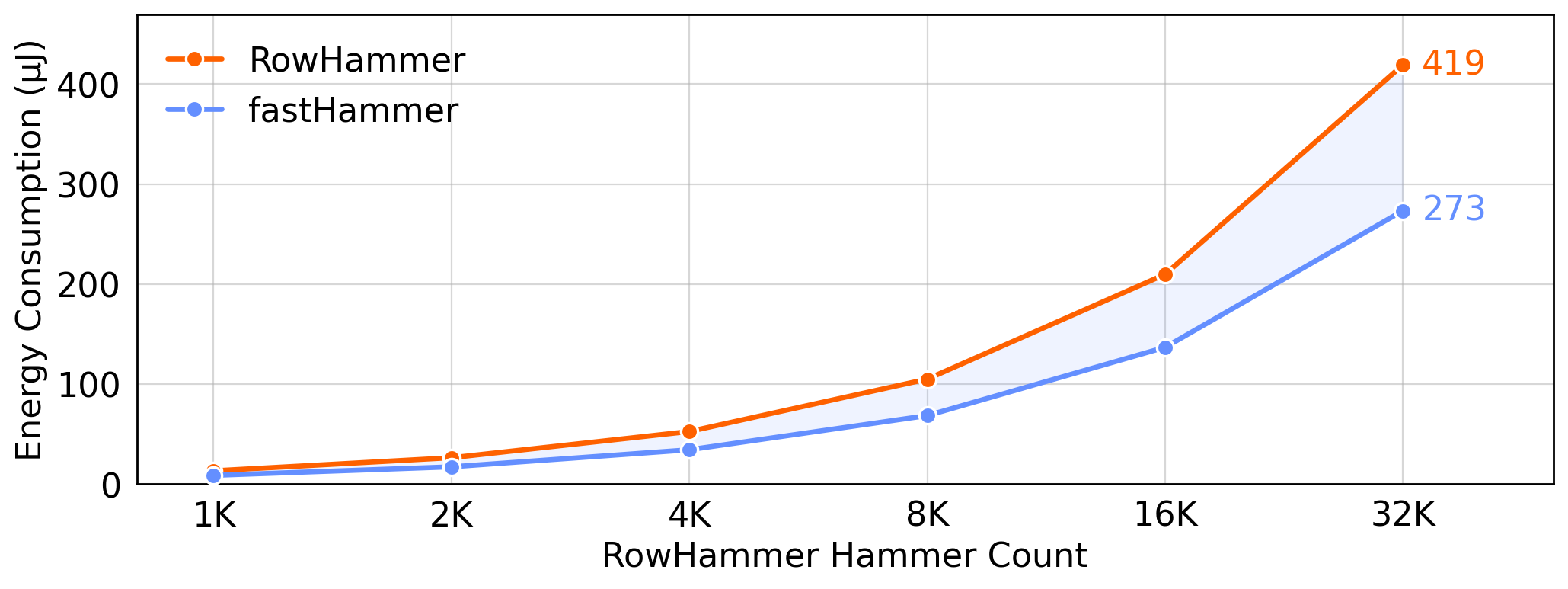}
\caption{Energy consumption of hammering a row with the \rh{} access pattern and
with \fastHammer{}.}
\label{fig:energy}
\end{figure}

\figref{fig:energy} shows the energy consumed to hammer a row with the conventional \rh{} access pattern and with \fastHammer{}. The x-axis shows the number of hammers performed with the \rh{} access pattern. The shaded region shows the energy that \fastHammer{} saves. 

We observe that \fastHammer{} consumes significantly less energy than the \rh{} pattern across all tested RDT values. At \param{32}K \rh{} hammers, \fastHammer{} consumes \param{273}~$\mu$J whereas \rh{} consumes \param{419}~$\mu$J, i.e., \fastHammer{} saves \param{146}~$\mu$J (\param{34.8\%}).
We conclude that \X{} reduces the energy cost of RDT profiling in addition to its latency.

\subsection{Latency Analysis}
We measure latency of profiling for a number of hammers with \fastHammer{} and \rh{} based on the three-step routine described in~\secref{sec:overview}. We follow the methodology described in~\cite{olgun2025variable} to tightly schedule DRAM commands based on the access patterns. For the hammer count of 4K, we observe that \fastHammer{} reduces the latency of profiling by \param{31.2\%} compared to \rh{} access pattern due to (i)~reduced timing constraints between \gls{pre} and \gls{act} commands and (ii)~reduction in hammer count (as shown in~\secref{sec:characterization}).
\section{Discussion}
\label{sec:discussion}
\head{\revc{Supporting \X{} with PRAC}}\labelc{2}\label{resp:c2}
\revc{PRAC requires a safe RDT configuration 
which \X{} can provide. To avoid triggering PRAC's preventive actions 
during profiling, the memory controller can use specialized precharge command that does \emph{not} update the 
PRAC counters, similar to prior work~\cite{vittal2025mopac}, or the system can employ \X{} in DRAM. }

\head{\revc{In-DRAM \X{}}} 
\revc{\X{} can run entirely in DRAM, if it supports autonomous maintenance operations (e.g., Self-Managing DRAM~\cite{hassan2024self}). Based on SMD, the DRAM chip can lock the region under
profiling, reject memory controller requests (e.g., via
\texttt{ACT\_NACK}), and perform \fastHammer{} internally. This way, DRAM chip can perform \fastHammer{} with fine-grained control over the \gls{pre}-to-\gls{act} latency and \X{} reads raw cell data, therefore avoiding on-die ECC masking bitflips.
DRAM can also prevent \rh{} mitigations to perform refreshes during profiling. After profiling, DRAM can use the profiled RDT directly in in-DRAM \rh{} mitigations or can report it to the memory
controller-based mitigations (e.g., via a mode register).}

\head{\revc{System Support for Reduced $t_{RP}$}}
\label{subsec:system_support}\label{resp:c3}
\labelc{3}\revc{\X{} requires the memory controller to apply a reduced $t_{RP}$
\emph{only} to profiling commands.
Modern memory controllers already expose
$t_{RP}$ as a programmable parameter (e.g., Intel's \texttt{TC\_PRE}
register~\cite{intel2020datasheet} and AMD's configurable
$t_{RP}$~\cite{amdryzenmaster}). Industry also proposes changing DRAM
timing parameters in-system to test installed DRAM beyond its rated
PRE-to-ACT timing~\cite{schoenborn2017memory} and to switch timing
profiles that include $t_{RP}$ at runtime without a
reboot~\cite{ley2024real}. Together, these capabilities provide the
support for reduced $t_{RP}$ in profiling.}
\section{Related Work}
To our knowledge, this is the first work to 1)~show, through a rigorous
characterization of \nCHIPS{} real DDR4 DRAM chips from \param{four} major
manufacturers, that intentionally exacerbating read disturbance with reduced DRAM
timing constraints preserves the read disturbance-weak cells and rows of the
conventional \rh{} access pattern, and 2)~exploit this property to design an
online RDT profiling technique that identifies a module's minimum RDT at low cost.

\noindent\textbf{DRAM Read Disturbance Characterization.} Many works~\cite{orosa2021deeper,kim2020revisiting,kim2014flipping, yaglikci2022understanding, lim2017active, park2016statistical, park2016experiments, ryu2017overcoming, yun2018study, lim2018study, luo2023rowpress, lang2023blaster, yaglikci2024spatial, nam2024dramscope,luo2025revisiting,tugrul2025understanding,yuksel2025pudhammer,he2023whistleblower,luo2024experimental,olgun2025variable} experimentally characterizes the real DRAM chip's read disturbance vulnerability. 
Two of these works reduce timing parameters to study read disturbance~\cite{tugrul2025understanding, yuksel2025pudhammer}, yet neither examines whether resulting bitflips preserve the read disturbance characteristics of the conventional \rh{} access pattern. We are the first to characterize the cell-level and row-level overlap between reduced-timing and \rh{} bitflips across \nMODULES{} real DRAM modules, and to show that this overlap enables fast RDT profiling.

\noindent\textbf{Retention Failure Profiling.} Prior works~\cite{liu2013experimental,qureshi2015avatar,patel2017reaper,
khan2014efficacy,khan2016parbor} propose methods to efficiently profile DRAM retention failures that are subject to the variable retention time. Similarly, read disturbance bitflips are subject to variable read disturbance~\cite{olgun2025variable}, yet no prior work provides an online RDT profiling technique. \X{} is the first online RDT profiling technique.

\section{Conclusion}
We present \X{}, an online RDT profiling technique that quickly profiles a DRAM module by intentionally exacerbating read disturbance. To do so, \X{} reduces the DRAM timing constraints during profiling, which both shortens the execution time of profiling and induces the first bitflip at a smaller activation count than conventional \rh{}.
Our characterization of \param{\nCHIPS{}} real DDR4 DRAM chips shows that \X{} measures a module's minimum RDT faster than the \rh{} profiler and 
\labela{3}\label{resp:a3}
\reva{preserves RDT ranking well enough to enable profiling a smaller candidate set to reprofile and refine its conservative minimum RDT measurement.}
Our end-to-end evaluations show that \X{} improves system performance over the \rh{} profiler. 
We hope that \X{} enables and inspires future work on fast and accurate online read disturbance profiling.

\section*{Acknowledgments}

{We thank the anonymous reviewers of HPCA 2026 for their feedback. We
thank the SAFARI Research Group members for their constructive feedback and for
providing a stimulating intellectual, scientific, scholarly environment. We
acknowledge the generous gift funding provided by our industrial partners
(especially Google, Huawei, Intel, Microsoft), which has been instrumental in
enabling the research we have been conducting on read disturbance in DRAM in
particular and memory systems in general. 
This work was in part supported by a Google Security and
Privacy Research Award and the Microsoft Swiss Joint Research Center.}

\balance
\bibliographystyle{unsrt}
\bibliography{refs}

\end{document}